\documentclass[prd,twocolumn,floatfix,amsmath,nofootinbib,amssymb,floatfix]{revtex4}
\usepackage{graphicx,color,dcolumn,booktabs,bm}
\usepackage{longtable,lscape}
\usepackage{pdfpages}
\usepackage{epsfig}
\usepackage{txfonts}
\usepackage{overpic}
\usepackage{amssymb}
\usepackage{makecell}
\usepackage{indentfirst}
\usepackage{float}
\usepackage{feynmf}   
\usepackage{slashed}  
\usepackage{cases}
\usepackage{color}
\usepackage{multirow}
\usepackage{threeparttable}
\usepackage{epstopdf}
\usepackage{enumerate}
\usepackage{subfigure}
\usepackage{diagbox}
\usepackage{graphicx,color,dcolumn,booktabs,bm}
\usepackage{mathrsfs}
\usepackage{cancel}
\usepackage[colorlinks,
            citecolor=blue,
            anchorcolor=red,
            menucolor=red,
            linkcolor=red,
            filecolor=red,
            runcolor=red,
            urlcolor=blue,
            frenchlinks=red]{hyperref}
\begin{document}
\title{The strong vertices of charmed mesons $D$, $D^{*}$ and charmonia $J/\psi$, $\eta_{c}$}
\author{Jie Lu$^{1,2}$}
\author{Guo-Liang Yu$^{1,2}$}
\email{yuguoliang2011@163.com}
\author{Zhi-Gang Wang$^{1}$}
\email{zgwang@aliyun.com}

\affiliation{$^1$ Department of Mathematics and Physics, North China
Electric Power University, Baoding 071003, People's Republic of
China\\$^2$ Hebei Key Laboratory of Physics and Energy Technology, North China Electric Power University, Baoding 071000, China}
\date{\today }

\begin{abstract}
In this work, the strong coupling constants of the vertices $DDJ/\psi$, $DD^{*}J/\psi$, $D^{*}D^{*}J/\psi$, $DD^{*}\eta_{c}$ and $D^{*}D^{*}\eta_{c}$ are calculated within the framework of the QCD sum rules. For each vertex, we analyze the momentum dependence of the coupling constants by considering all possible off-shell cases. In these analyses, we consider the contributions of the vacuum condensate terms $\langle\overline{q}q\rangle$, $\langle\overline{q}g_{s}\sigma Gq\rangle$, $\langle g_{s}^{2}G^{2}\rangle$, $\langle f^{3}G^{3}\rangle$ and $\langle\overline{q}q\rangle\langle g_{s}^{2}G^{2}\rangle$. Then, the momentum dependent coupling constants are fitted into analytical functions $g(Q^2)$ and are extrapolated into time-like regions. The values of strong coupling constants are obtained by using on-shell conditions of the intermediate mesons($Q^2=-m^2$). The final results are as follows, $g_{DDJ/\psi}=5.01^{+0.58}_{-0.16}$, $g_{DD^{*}J/\psi}=3.55^{+0.20}_{-0.21}$GeV$^{-1}$, $g_{D^{*}D^{*}J/\psi}=5.10^{+0.59}_{-0.43}$, $g_{DD^{*}\eta_{c}}=3.68^{+0.39}_{-0.11}$ and $g_{D^{*}D^{*}\eta_{c}}=4.87^{+0.42}_{-0.40}$GeV$^{-1}$.
\end{abstract}


\maketitle

\section{Introduction}\label{sec1}

\begin{figure}[htbp]
\centering
\includegraphics[width=8cm]{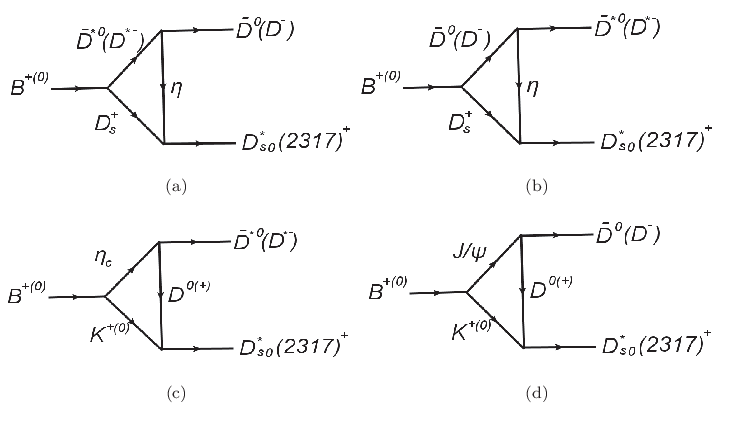}
\caption{Triangle diagrams accounting for the $B$ decays: (a)$B^{+(0)}\to D^{+}_{s}\bar{D}^{*0}(D^{*-})\to D_{s0}^{*}(2317)^{+}\bar{D}^{0}(D^{-})$, (b)$B^{+(0)}\to D^{+}_{s}\bar{D}^{0}(D^{-})\to D_{s0}^{*}(2317)^{+}\bar{D}^{*0}(D^{*-})$, (c)$B^{+(0)}\to \eta_{c}K^{+(0)}\to D_{s0}^{*}(2317)^{+}\bar{D}^{*0}(D^{*-})$ and (d)$B^{+(0)}\to J/\psi K^{+0}\to D_{s0}^{*}(2317)^{+}\bar{D}^{0}(D^{-})$.}
\label{TD1}
\end{figure}

\begin{figure}[htbp]
\centering
\includegraphics[width=8cm]{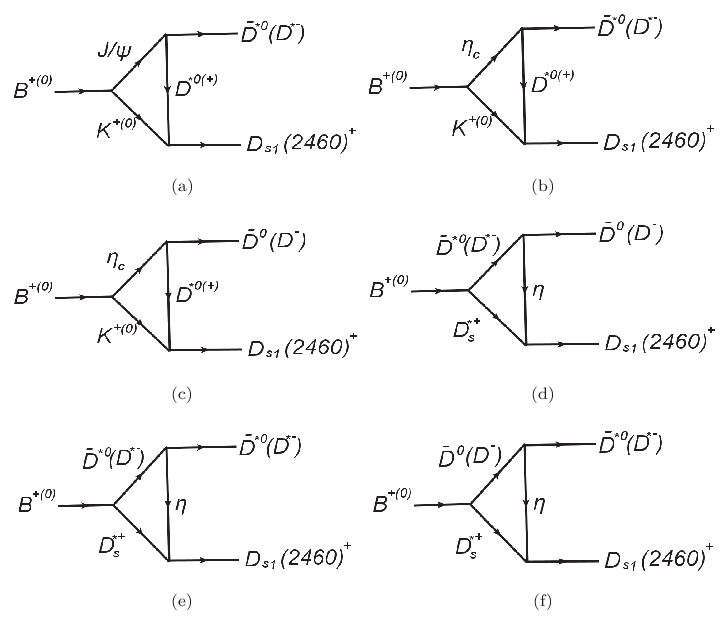}
\caption{Triangle diagrams accounting for the $B$ decays: (a)$B^{+(0)}\to J/\psi K^{+(0)} \to D_{s1}(2460)^{+}\bar{D}^{*0}(D^{*-})$, (b)$B^{+(0)}\to \eta_{c}K^{+(0)} \to D_{s1}(2460)^{+}\bar{D}^{*0}(D^{*-})$, (c)$B^{+(0)}\to \eta_{c}K^{+(0)}\to D_{s1}(2460)\bar{D}^{0}(D^{-})$, (d)$B^{+(0)}\to D_{s}^{*+}\bar{D}^{*-}(D^{*-}) \to D_{s1}(2460)^{+}\bar{D}^{0}(D^{-})$, (e)$B^{+(0)}\to D_{s}^{*+}\bar{D}^{*-}(D^{*-}) \to D_{s1}(2460)^{+}\bar{D}^{*0}(D^{*-})$ and (f)$B^{+(0)}\to D_{s}^{*+}\bar{D}^{0}(D^{-}) \to D_{s1}(2460)^{+}\bar{D}^{*0}(D^{*-})$.}
\label{TD2}
\end{figure}

In recent years, many exotic hadrons which are beyond the usual quark-model were observed in experments\cite{BaBar:2003oey,BaBar:2005hhc,Belle:2004lle,BaBar:2006gsq,CDF:2009jgo,Belle:2009rkh,Belle:2011aa,Xiao:2013iha,LHCb:2014zfx,Belle:2014nuw,LHCb:2015yax,CDF:2011pep,LHCb:2017iph}. Some of them have exotic quantum numbers and are interpreted as tetraquark states, pentaquark states, hadron molecular states, quark-gluon hybrids, glueballs and many others\cite{Branz:2009yt,Mahajan:2009pj,Liu:2009iw,Liu:2009qhy,Dias:2013xfa,Wang:2014gwa,Wang:2013zra,Zhao:2011sd,Brodsky:2014xia}. The inner structure of these exotic hadrons can not be determined only by the mass spectrum. We need to further study their production processes or decay behaviours, where the coupling constants become particularly important. For example, the exotic states $D_{s0}^{*}$(2317) and $D_{s1}$(2460) were discovered in the decay processes $B\to\bar{D}^{(*)}D_{s0}^{*}\mathrm{(2317)}$ and $B\to\bar{D}^{(*)}D_{s1}\mathrm{(2460)}$\cite{BaBar:2003oey,CLEO:2003ggt,Belle:2003guh}, and their decay  branching fractions were also measured in the experiment\cite{ParticleDataGroup:2022pth}.
If they are interpreted as $D^{(*)}K$ or $D_{s}^{(*)}\eta$ molecular states, their production processes can be studied in the triangle mechanism\cite{Liu:2022dmm}, which are shown in Figs. \ref{TD1} and \ref{TD2}. In these processes, the three meson vertices $DD^{*}\eta$, $D^{*}D^{*}\eta$, $DDJ/\psi$, $D^{*}D^{*}J/\psi$, $DD^{*}\eta_{c}$ and $D^{*}D^{*}\eta_{c}$ are especially important for us to analyze their production processes.

The QCD is perturbative field theory which has been successfully applied in the high energy region. However, the strong coupling constant between the hadrons lies in the low energy region, which can not be studied by perturbative method. As for the mesons composed of u, d or s quark, the couplings between these mesons have been constrained by the SU(3) chiral symmetry and phenomenological analyses of low energy reactions\cite{deSwart:1963pdg,Holzenkamp:1989tq,Carvalho:1999he}. For the charmed mesons, we can use SU(4) chiral symmetry and phenomenological analysis to get the interaction Lagrangion\cite{Matinyan:1998cb,Lin:1999ad,Oh:2000qr}. Because the mass of c quark is much heavier than u, d and s quark, the SU(4) chiral symmetry is badly broken. Therefore, the accurate calculation of the strong coupling constants is of great significance to the study of the destruction of SU(4) chiral symmetry.


Since the strong interaction is non-perturbative in the low energy region, it is difficult to calculate the coupling constants from the QCD first principle. Except for lattice calculation\cite{Altmeyer:1995qx}, some phenomenological methods are usually employed to carry out this work such as the QCDSR\cite{Navarra:2000ji,Navarra:2001ju,RodriguesdaSilva:2003hh,Aydin:2004ty,Bracco:2004rx,Aydin:2006ck,Bracco:2006xf,Bracco:2007sg,Bracco:2010bf,OsorioRodrigues:2010fen,Azizi:2010jj,Sundu:2011vz,Cerqueira:2011za,Cui:2011zq,Cui:2012wk,Bracco:2011pg,Wang:2013iia,Yu:2015xwa,Rodrigues:2017qsm}, LCSR\cite{Colangelo:1995ph,Aliev:1996bp,Colangelo:1997rp,Dai:1998ve,Zhu:1998vf,Khodjamirian:1999hb,Li:2002pp,Kim:2001es,Wang:2006bs,Wang:2006ida,Wang:2007mc,Wang:2007zm,Wang:2008tm,Wang:2007ci,Li:2007dv,Khodjamirian:2020mlb}, VMD model\cite{Lin:1999ad,Oh:2000qr} and other methods\cite{Li:2002pp,Deandrea:2003pv}. The QCDSR\cite{Shifman:1978by} is one of the most powerful non-perturbative methods to analyze the strong vertices. In our previous work, we have analyzed the vertices $D_{s}^{*}D_{s}\phi$, $D_{2}^{*}D^{*}\pi$, $D_{s2}^{*}D^{*}K$, $D_{2}^{*}D\rho$, $D_{2}^{*}D\omega$, $D_{s2}^{*}D_{s}\phi$ using the three-point correlation\cite{Yu:2015xwa,Yu:2019sqp,Li:2015xka}. As a continuation of these tasks, we systematically analyze the strong vertices of the charmed mesons $D$, $D^{*}$ and the charmonia $J/\psi$, $\eta_{c}$ using the three-point QCDSR. Although some strong vertices $DDJ/\psi$, $DD^{*}J/\psi$, $D^{*}D^{*}J/\psi$ and $DD^{*}\eta_{c}$ were already analyzed by other collaborations in QCDSR\cite{Bracco:2011pg,Rodrigues:2017qsm}. However, the high dimension condensate terms such as the $\langle\overline{q}g_{s}\sigma Gq\rangle$, $\langle f^{3}G^{3}\rangle$ and $\langle\bar{q}q\rangle\langle g_{s}^{2}G^{2}\rangle$ were neglected in their studies, which is significant to the accuracy of the final results.

This article is organized as follows. After the introduction in Sec. \ref{sec1}, we analyze the strong vertices $DDJ/\psi$, $DD^{*}J/\psi$, $D^{*}D^{*}J/\psi$, $DD^{*}\eta_{c}$ and $D^{*}D^{*}\eta_{c}$ with the QCDSR in Sec. \ref{sec2}. In these analyses, all off-shell cases of the intermediate mesons are considered. In the QCD side, we consider the perturbative contribution and vacuum condensate terms $\langle\overline{q}q\rangle$, $\langle\overline{q}g_{s}\sigma Gq\rangle$, $\langle g_{s}^{2}G^{2}\rangle$, $\langle f^{3}G^{3}\rangle$ and $\langle\overline{q}q\rangle\langle g_{s}^{2}G^{2}\rangle$. In Sec. \ref{sec3}, we present the numerical results and discussions. Sec. \ref{sec4} is reserved for our conclusions. Some important figures and formulas will be shown in the Appendixes A-C.

\section{QCD sum rules} \label{sec2}

Firstly, we write down the three-point correlation function,
\begin{eqnarray}
\notag
\Pi (p,p') &&= {i^2}\int {{d^4}x} {d^4}y{e^{ip'x}}{e^{i(p - p')y}}\\
&&\times \left\langle 0 \right.| T\{ {J_C}(x){J_B}(y)J_A^ \dagger (0)\} |\left. 0 \right\rangle
\end{eqnarray}
where $T$ denotes the time ordered product and $J_{A}^{\dagger}, J_{B}$ and $J_{C}$ are the meson interpolating currents. The subscripts A, B, and C denote three mesons in each vertex, where B is the intermediate meson and is off-shell. Assignments of the mesons for each vertex are presented in Table~\ref{OS}.

\begin{table}[htbp]
\caption{The assignments of the mesons A, B and C for each vertex where B denotes the off-shell mesons.}
\label{OS}
\begin{tabular}{p{1.8cm}<{\centering} p{1.8cm}<{\centering} p{1.8cm}<{\centering} p{1.8cm}<{\centering} }
\hline
\hline
Vertices&A&B(off-shell)&C \\ \hline
\multirow{2}*{$DDJ/\psi$}&$D$&$J/\psi$&$D$  \\
~&$D$&$D$&$J/\psi$   \\  \hline
\multirow{3}*{$DD^{*}J/\psi$}&$D^{*}$&$J/\psi$&$D$  \\
~&$D^{*}$&$D$&$J/\psi$   \\
~&$D$&$D^{*}$&$J/\psi$   \\ \hline
\multirow{2}*{$D^{*}D^{*}J/\psi$}&$D^{*}$&$J/\psi$&$D^{*}$  \\
~&$D^{*}$&$D^{*}$&$J/\psi$   \\ \hline
\multirow{3}*{$DD^{*}\eta_{c}$}&$D$&$\eta_{c}$&$D^{*}$  \\
~&$D^{*}$&$D$&$\eta_{c}$   \\
~&$D$&$D^{*}$&$\eta_{c}$  \\ \hline
\multirow{2}*{$D^{*}D^{*}\eta_{c}$}&$D^{*}$&$\eta_{c}$&$D^{*}$  \\
~&$D^{*}$&$D^{*}$&$\eta_{c}$ \\ \hline\hline
\end{tabular}
\end{table}

The meson interpolating currents in these analyses are as follows,
\begin{eqnarray}
\notag
{J_D}(x) = \bar u(x)i{\gamma _5}c(x)\\
\notag
{J_{{D^*}}}(x) = \bar u(x){\gamma _\mu }c(x)\\
\notag
{J_{J/\psi }}(x) = \bar c(x){\gamma _\mu }c(x)\\
{J_{{\eta _c}}}(x) = \bar c(x)i{\gamma _5}c(x)
\end{eqnarray}
In the framework of QCD sum rules, the correlation function is firstly calculated in two levels: the hadron level which is called the phenomenological side, and the quark level which is called the QCD side. Then we use the quark hadron duality coordinate the calculation of these two levels. In the following section, we will obtain the sum rules about these strong vertices.

\subsection{The Phenomenological side}\label{2.1}

In phenomenological side, we insert complete sets of hadronic states with the same quantum numbers as the currents $J_{A}^{\dagger}$, $J_{B}$ and $J_{C}$ into the correlation function. Using the dispersion relation\cite{Bracco:2011pg}, the correlation function can be written as,
\begin{eqnarray}\label{eq:3}
\notag
\Pi (p,p') &&= \frac{{\left\langle 0 \right.|{J_C}(0)|\left. {C(p')} \right\rangle \left\langle 0 \right.|{J_B}(0)|\left. {B(q)} \right\rangle }}{{(m_A^2 - {p^2})(m_C^2 - p{'^2})(m_B^2 - {q^2})}}\\
&&\times \left\langle {A(p)} \right.|J_A^ \dagger (0)|\left. 0 \right\rangle \left\langle {B(q)C(p')|\left. {A(p)} \right\rangle } \right. + h.c.
\end{eqnarray}
where $h.c.$ represents the contributions of higher resonances and continuum states of each meson. The meson vacuum matrix elements appearing in this equation are substituted by the following parameterized equations,
\begin{eqnarray} \label{eq:4}
\notag
&&\langle0|J_{D}(0)|D\rangle =\frac{f_{D}m_{D}^{2}}{m_{c}}\\
\notag
&&\langle0|J_{D^{*}}(0)|D^{*}\rangle=f_{D^{*}}m_{D^{*}}\zeta _{\mu }\\
\notag
&&\langle0|J_{J/\psi }(0)|J/\psi\rangle=f_{J/\psi}m_{J/\psi}\xi_{\mu}\\
&&\langle0|J_{\eta _{c}}(0)|\eta _{c}\rangle=\frac{f_{\eta_{c}}m_{\eta_{c}}^{2}}{2m_{c}}
\end{eqnarray}
where $f_{D}$, $f_{D^{*}}$, $f_{J/\psi}$ and $f_{\eta_{c}}$ are the meson decay constants, $\zeta_{\mu}$ and $\xi_{\mu}$ are the polarization vectors of $D^{*}$ and $J/\psi$, respectively. All of the meson vertex matrix elements in Eq. (\ref{eq:3}) can be obtained by the following effective Lagrangian \cite{Xiao:2019mvs,Wu:2021ezz},
\begin{eqnarray} \label{eq:5}
\notag
&&{\mathscr{L}_{DDJ/\psi}} = i{g_{DDJ/\psi }}{\psi _\alpha }({\partial ^\alpha }D\bar D  - D{\partial ^\alpha }\bar D ) \\
\notag
&&{\mathscr{L}_{DD^{*}J/\psi}} =  - {g_{{D^*}DJ/\psi }}{\varepsilon ^{\alpha \beta \rho \tau }}{\partial _\alpha }{\psi _\beta }({\partial _\rho }D_\tau ^*\bar D  + D{\partial _\rho }\bar D _\tau ^*) \\
\notag
&&{\mathscr{L}_{D^{*}D^{*}J/\psi}} = i{g_{{D^*}{D^*}J/\psi }}[{\psi ^\alpha }({\partial _\alpha }{D^{*\beta }}\bar {D_\beta ^*}  - {D^{*\beta }}{\partial _\alpha }\bar {D_\beta ^*} ) \\
\notag
&&+ ({\partial _\alpha }{\psi _\beta }{D^{*\beta }} - {\psi _\beta }{\partial _\alpha }{D^{*\beta }}){\bar D ^{*\alpha }} + {D^{*\alpha }}({\psi ^\beta }{\partial _\alpha }\bar {D_\beta ^*}  - {\partial _\alpha }{\psi _\beta }{\bar D ^{*\beta }})]  \\
\notag
&&{\mathscr{L}_{DD^{*}\eta_{c}}} = i{g_{{D^*}D{\eta _c}}}[{D^{*\alpha }}({\partial _\alpha }{\eta _c}\bar D  - {\eta _c}{\partial _\alpha }\bar D ) \\
\notag
&& + ({\partial _\alpha }{\eta _c}D - {\eta _c}{\partial _\alpha }D){\bar D ^{*\alpha }}]  \\
&&{\mathscr{L}_{D^{*}D^{*}\eta_{c}}} =  - {g_{{D^*}{D^*}{\eta _c}}}{\varepsilon ^{\alpha \beta \rho \tau }}{\partial _\alpha }D_\beta ^*{\partial _\rho }\bar D_\tau ^*{\eta _c}
\end{eqnarray}
Expressions of the meson vertex elements for all vertices are as follows,
\begin{eqnarray} \label{eq:6}
\notag
&&\left\langle {D(p')J/\psi (q)|\left. {D(p)} \right\rangle } \right. = g_{DDJ/\psi }^{J/\psi }({q^2})\xi _\alpha ^*{(p + p')^\alpha }\\
\notag
&&\left\langle {D(q)J/\psi (p')|\left. {D(p)} \right\rangle } \right. = g_{DDJ/\psi }^D({q^2}){\xi _\alpha ^* }{(p + q)^\alpha }\\
\notag
&&\left\langle {D(p')J/\psi (q)|\left. {{D^*}(p)} \right\rangle } \right. =  - g_{D{D^*}J/\psi }^{J/\psi }({q^2}){\varepsilon ^{\alpha \beta \rho \tau }}{\xi _\alpha }{\zeta _\beta }{p_\rho }p{'_\tau }\\
\notag
&&\left\langle {D(q)J/\psi (p')|\left. {{D^*}(p)} \right\rangle } \right. =  - g_{D{D^*}J/\psi }^D({q^2}){\varepsilon ^{\alpha \beta \rho \tau }}{\xi _\alpha }{\zeta _\beta }{p_\rho }p{'_\tau }\\
\notag
&&\left\langle {{D^*}(q)J/\psi (p')|\left. {D(p)} \right\rangle } \right. =  g_{D{D^*}J/\psi }^{{D^*}}({q^2}){\varepsilon ^{\alpha \beta \rho \tau }}{\xi _\alpha }{\zeta _\beta }p{'_\rho }{p_\tau }\\
\notag
&&\left\langle {{D^*}(p')J/\psi (q)|\left. {{D^*}(p)} \right\rangle } \right. = g_{{D^*}{D^*}J/\psi }^{J/\psi }[{(p + p')^\alpha }{\xi ^{*}_{\alpha }}{\zeta ^{'\beta} }{\zeta _\beta ^*}  \\
\notag
&&- {(p + q)^\alpha }{\zeta ^{'*}_{\alpha }}\xi _\beta ^*{\zeta ^\beta } - {(p' + q)^\alpha }{\zeta _\alpha }{\xi ^{*\beta }}\zeta _\beta ^{'*}]\\
\notag
&&\left\langle {{D^*}(q)J/\psi (p')|\left. {{D^*}(p)} \right\rangle } \right. = g_{{D^*}{D^*}J/\psi }^{{D^*}}[{(p + q)^\alpha }{\xi ^{*}_{\alpha }}{\zeta ^\beta }\zeta _\beta ^{'*} \\
\notag
&&- {(p + p')^\alpha }{\zeta ^{'*}_{\alpha }}\xi _\beta ^*{\zeta ^\beta } - {(q + p')^\alpha }{\zeta _\alpha }{\xi ^{*\beta }}\zeta _\beta ^{'*}]\\
\notag
&&\left\langle {D(p'){\eta _c}(q)|\left. {{D^*}(p)} \right\rangle } \right. =  - g_{D{D^*}{\eta _c}}^{{\eta _c}}({q^2}){\zeta _\alpha }{(q - p')^\alpha }\\
\notag
&&\left\langle {D(q){\eta _c}(p')|\left. {{D^*}(p)} \right\rangle } \right. =  - g_{D{D^*}{\eta _c}}^D({q^2}){\zeta _\alpha }{(p' - q)^\alpha }\\
\notag
&&\left\langle {{D^*}(q){\eta _c}(p')|\left. {D(p)} \right\rangle } \right. =  - g_{D{D^*}{\eta _c}}^{{D^*}}({q^2}){\zeta ^{*}_{\alpha }}{(p + p')^\alpha }\\
\notag
&&\left\langle {{D^*}(p'){\eta _c}(q)|\left. {{D^*}(p)} \right\rangle } \right. =  g_{{D^*}{D^*}{\eta _c}}^{{\eta _c}}({q^2}){\varepsilon ^{\alpha \beta \rho \tau }}{\zeta _\alpha }\zeta _\beta ^{'*}{p_\rho }p{'_\tau }\\
&&\left\langle {{D^*}(q){\eta _c}(p')|\left. {{D^*}(p)} \right\rangle } \right. =  g_{{D^*}{D^*}{\eta _c}}^{{D^*}}({q^2}){\varepsilon ^{\alpha \beta \rho \tau }}{\zeta ^{'} _\alpha }\zeta _\beta ^*{p_\rho }{q_\tau }
\end{eqnarray}
where $\xi _{\alpha}$ and $\zeta^{(')}_{\alpha}$ are the polarization vectors of $J/\psi$ and $D^{*}$ respectively, $\varepsilon^{\alpha\beta\rho\tau}$ is the Levi-Civita tensor, and $q=p-p'$. The superscripts of $g$ in these equations denote the intermediate mesons which are off-shell, and subscripts refer to the type of vertices. For example, $g_{DDJ/\psi }^{J/\psi }$ represents the strong coupling constant of the vertex $DDJ/\psi$, where $J/\psi$ is the intermediate meson. From Eqs. (\ref{eq:4})-(\ref{eq:6}), the correlation functions in hadron side will be obtained, and it can be expanded into different tensor structures. Theoretically, each structure can be used to carry out the calculation and can lead to the same result. However, the results obtained from different tensor structures have different uncertainties which originate from the truncation of OPE in the QCD side and from different contributions of the continuum\cite{Bracco:1999xe}. In order to choose appropriate tensor structures and obtain reliable results, we adopt the traditional way to solve this problem, where the conditions of pole dominance(pole contributions should be larger than $40\%$) and appearance of Borel window should be satisfied.

\subsection{The QCD side}\label{sec2.2}

In QCD side, we do the operator product expansion(OPE) of the correlation function by contracting the quark fields with Wick's theorem. The correlation functions in QCD side for vertices $DDJ/\psi$, $DD^{*}J/\psi$, $D^{*}D^{*}J/\psi$, $DD^{*}\eta_{c}$ and $D^{*}D^{*}\eta_{c}$ are expressed as follows,
\begin{eqnarray}\label{eq:7}
\notag
\Pi _\mu ^{J/\psi }(p,p') &&= \int {{d^4}x{d^4}y{e^{ip'x}}{e^{i(p - p')y}}} \\
\notag
&&\times Tr\{ {C^{nk}}(y){\gamma _5}{U^{km}}( - x){\gamma _5}{C^{mn}}(x - y){\gamma _\mu }\} \\
\notag
\Pi _\mu ^D(p,p') &&= \int {{d^4}x{d^4}y{e^{ip'x}}{e^{i(p - p')y}}} \\
&&\times Tr\{ {\gamma _\mu }{C^{nk}}(x){\gamma _5}{U^{km}}( - y){\gamma _5}{C^{mn}}(y - x)\}
\end{eqnarray}
\begin{eqnarray}\label{eq:8}
\notag
\Pi _{\mu \nu }^{J/\psi }(p,p') &&=  - i\int {{d^4}x{d^4}y{e^{ip'x}}{e^{i(p - p')y}}}\\
\notag
&&\times Tr\{ {C^{nk}}(y){\gamma _\nu }{U^{km}}( - x){\gamma _5}{C^{mn}}(x - y){\gamma _\mu }\} \\
\notag
\Pi _{\mu \nu }^D(p,p') &&=  - i\int {{d^4}x{d^4}y{e^{ip'x}}{e^{i(p - p')y}}} \\
\notag
&&\times Tr\{ {\gamma _\mu }{C^{nk}}(x){\gamma _\nu }{U^{km}}( - y){\gamma _5}{C^{mn}}(y - x)\}  \\
\notag
\Pi _{\mu \nu }^{{D^*}}(p,p') &&=  - i\int {{d^4}x{d^4}y{e^{ip'x}}{e^{i(p - p')y}}} \\
&&\times Tr\{ {\gamma _\mu }{C^{nk}}(x){\gamma _5}{U^{km}}( - y){\gamma _\nu }{C^{mn}}(y - x)\}
\end{eqnarray}
\begin{eqnarray}\label{eq:9}
\notag
\Pi _{\mu \nu \sigma }^{J/\psi }(p,p') &&= \int {{d^4}x{d^4}y{e^{ip'x}}{e^{i(p - p')y}}}  \\
\notag
&&\times Tr\{ {C^{nk}}(y){\gamma _\nu }{U^{km}}( - x){\gamma _\sigma }{C^{mn}}(x - y){\gamma _\mu } \} \\
\notag
\Pi _{\mu \nu \sigma }^{{D^*}}(p,p') &&= \int {{d^4}x{d^4}y{e^{ip'x}}{e^{i(p - p')y}}} \\
&&\times Tr\{ {\gamma _\mu }{C^{nk}}(x){\gamma _\sigma }{U^{km}}( - y){\gamma _\nu }{C^{mn}}(y - x)\}
\end{eqnarray}
\begin{eqnarray}\label{eq:10}
\notag
\Pi _\mu ^{{\eta _c}}(p,p') &&= \int {{d^4}x{d^4}y{e^{ip'x}}{e^{i(p - p')y}}} \\
\notag
&&\times Tr\{ {C^{nk}}(y){\gamma _\mu }{U^{km}}( - x){\gamma _5}{C^{mn}}(x - y){\gamma _5}\} \\
\notag
\Pi _\mu ^D(p,p') &&= \int {{d^4}x{d^4}y{e^{ip'x}}{e^{i(p - p')y}}} \\
\notag
&&\times Tr\{ {\gamma _5}{C^{nk}}(x){\gamma _\mu }{U^{km}}( - y){\gamma _5}{C^{mn}}(y - x)\}  \\
\notag
\Pi _\mu ^{{D^*}}(p,p') &&= \int {{d^4}x{d^4}y{e^{ip'x}}{e^{i(p - p')y}}}  \\
&&\times Tr\{ {\gamma _5}{C^{nk}}(x){\gamma _5}{U^{km}}( - y){\gamma _\mu }{C^{mn}}(y - x)\}
\end{eqnarray}
\begin{eqnarray}\label{eq:11}
\notag
\Pi _{\mu \nu }^{{\eta _c}}(p,p') &&=  - i\int {{d^4}x{d^4}y{e^{ip'x}}{e^{i(p - p')y}}} \\
\notag
&&\times Tr\{ {C^{nk}}(y){\gamma _\nu }{U^{km}}( - x){\gamma _\mu }{C^{mn}}(x - y){\gamma _5}\}  \\
\notag
\Pi _{\mu \nu }^{{D^*}}(p,p') &&=  - i\int {{d^4}x{d^4}y{e^{ip'x}}{e^{i(p - p')y}}} \\
&&\times Tr\{ {\gamma _5}{C^{nk}}(x){\gamma _\mu }{U^{km}}( - y){\gamma _\nu }{C^{mn}}(y - x)\}
\end{eqnarray}
It is the same as $g$ in Eqs. (\ref{eq:4})-(\ref{eq:6}), the superscripts of $\Pi$ in these above equations denote the intermediate mesons which are off-shell. $U^{ij}(x)$ and $C^{ij}(x)$ are the full propagators of $u(d)$ and $c$ quarks which can be written as\cite{Wang:2014yza},
\begin{eqnarray}
\notag
{U^{ij}}(x) &&= \frac{{i{\delta ^{ij}}\slashed{x}}}{{2{\pi ^2}{x^4}}} - \frac{{{\delta ^{ij}}{m_q}}}{{4{\pi ^2}{x^4}}} - \frac{{{\delta ^{ij}}\left\langle {\bar qq} \right\rangle }}{{12}} + \frac{{i{\delta ^{ij}}\slashed{x}{m_q}\left\langle {\bar qq} \right\rangle }}{{48}}\\
\notag
&& - \frac{{{\delta ^{ij}}{x^2}\left\langle {\bar q{g_s}\sigma Gq} \right\rangle }}{{192}} + \frac{{i{\delta ^{ij}}{x^2}\slashed{x}{m_q}\left\langle {\bar q{g_s}\sigma Gq} \right\rangle }}{{1152}}\\
\notag
&& - \frac{{i{g_s}G_{\alpha \beta }^at_{ij}^a(\slashed{x}{\sigma ^{\alpha \beta }} + {\sigma ^{\alpha \beta }}\slashed x)}}{{32{\pi ^2}{x^2}}} - \frac{{i{\delta ^{ij}}{x^2}\slashed{x}g_s^2{{\left\langle {\bar qq} \right\rangle }^2}}}{{7776}}\\
\notag
&& - \frac{{{\delta ^{ij}}{x^4}\left\langle {\bar qq} \right\rangle \left\langle {g_s^2GG} \right\rangle }}{{27648}} - \frac{{\left\langle {{{\bar q}^j}{\sigma ^{\mu \nu }}{q^i}} \right\rangle {\sigma _{\mu \nu }}}}{8}\\
\notag
&& - \frac{{\left\langle {{{\bar q}^j}{\gamma ^\mu }{q^i}} \right\rangle {\gamma _\mu }}}{4} + ...\\
\notag
{C^{ij}}(x) &&= \frac{i}{{{{(2\pi )}^4}}}\int {{d^4}k} {e^{ - ik \cdot x}}\{ \frac{{{\delta ^{ij}}}}{{\slashed k - {m_c}}}\\
\notag
&& - \frac{{{g_s}G_{\alpha \beta }^nt_{ij}^n}}{4}\frac{{{\sigma ^{\alpha \beta }}(\slashed{k} + {m_c}) + (\slashed{k} + {m_c}){\sigma ^{\alpha \beta }}}}{{{{({k^2} - m_c^2)}^2}}}\\
\notag
&& + \frac{{{g_s}{D_\alpha }G_{\beta \lambda }^nt_{ij}^n({f^{\lambda \beta \alpha }} + {f^{\lambda \alpha \beta }})}}{{3{{({k^2} - m_c^2)}^4}}}\\
\notag
&& - \frac{{g_s^2{{({t^a}{t^b})}_{ij}}G_{\alpha \beta }^aG_{\mu \nu }^b({f^{\alpha \beta \mu \nu }} + {f^{\alpha \mu \beta \nu }} + {f^{\alpha \mu \nu \beta }})}}{{4{{({k^2} - m_c^2)}^5}}} \\
&&+ ...\}
\end{eqnarray}
where $\langle g_{s}^{2}G^{2}\rangle=\langle g_{s}^{2}G^{n}_{\alpha\beta}G^{n\alpha\beta}\rangle$, $D_{\alpha}=\partial_{\alpha}-ig_{s}G^{n}_{\alpha}t^{n}$, $t^{n}=\frac{\lambda^{n}}{2}$, $\lambda^{n}$ is the Gell-Mann matrix,  $i$, $j$ are color indices, $q=u(d)$, $\sigma_{\alpha\beta}=\frac{i}{2}[\gamma_{\alpha},\gamma_{\beta}]$ and $f^{\lambda\alpha\beta}$, $f^{\alpha\beta\mu\nu}$ have the following forms,
\begin{eqnarray}
{f^{\lambda \alpha \beta }} = (\slashed k + {m_c}){\gamma ^\lambda }(\slashed k + {m_c}){\gamma ^\alpha }(\slashed k + {m_c}){\gamma ^\beta }(\slashed k + {m_c})
\end{eqnarray}
\begin{eqnarray}
\notag
{f^{\alpha \beta \mu \nu }} = && (\slashed k + {m_c}){\gamma ^\alpha }(\slashed k + {m_c}){\gamma ^\beta }(\slashed k + {m_c})\\
&&{\gamma ^\mu }(\slashed k + {m_c}){\gamma ^\nu }(\slashed k + {m_c})
\end{eqnarray}

The correlation functions of $DDJ/\psi$ and $DD^{*}\eta_{c}$ in Eqs. (\ref{eq:7}) and (\ref{eq:10}) have the same tensor structures, which can be expanded into the following structures,
\begin{eqnarray}
\Pi_{\mu}(p,p')=\Pi_{1}(p^{2},p'^{2},q^{2})p_{\mu}+ \Pi_{2}(p^{2},p'^{2},q^{2})p'_{\mu}
\end{eqnarray}
We choose the scalar amplitude $\Pi_{1}^{J/\psi(D)}$ and $\Pi_{1}^{D^{*}}$ to obtain $g_{DDJ/\psi}^{J/\psi(D)}$ and $g_{DD^{*}\eta_{c}}^{D^{*}}$, and use $\Pi_{2}^{\eta_{c}(D)}$ to obtain $g_{DD^{*}\eta_{c}}^{\eta_{c}(D)}$.

For the vertices $DD^{*}J/\psi$ and $D^{*}D^{*}\eta_{c}$, their correlation functions in Eqs. (\ref{eq:8}) and (\ref{eq:11}) have only one structure,
\begin{eqnarray}
\Pi_{\mu\nu}(p,p')= \Pi(p^{2},p'^{2},q^{2})\varepsilon_{\mu\nu\alpha\beta}p^{\alpha}p'^{\beta}
\end{eqnarray}
where $\varepsilon_{\mu\nu\alpha\beta}$ is the 4-dimension Levi-Civita tensor. It is natural that this above structure will be used to analyze coupling constants $g_{DD^{*}J/\psi}^{J/\psi(D,D^{*})}$ and $g_{D^{*}D^{*}\eta_{c}}^{\eta_{c}(D^{*})}$.

There are three Lorentz indices for the correlation functions of vertex $D^{*}D^{*}J/\psi$ in Eq. (\ref{eq:9}), which will lead to complicated tensor structures. Using the metric tensor $g^{\mu\nu}$, these correlation functions can be reduced as,
\begin{eqnarray}
\notag
&&g^{\mu\nu}\Pi_{\mu\nu\sigma}(p,p') = \slashed{\Pi}_{\sigma}(p,p') \\
&&= \slashed{\Pi}_{1}(p^{2},p'^{2},q^{2})p_{\sigma}+\slashed{\Pi}_{2}(p^{2},p'^{2},q^{2})p'_{\sigma}
\end{eqnarray}
where $\slashed{\Pi}_{\sigma}$ is named as reduced correlation function. In the following analysis, $\slashed{\Pi}_{1}^{J/\psi(D^{*})}$ will be used to obtain  $g_{D^{*}D^{*}J/\psi}^{J/\psi(D^{*})}$.

In the QCD side, we use $\Pi^{\mathrm{OPE}}$ to represent the selected scalar invariant amplitude which is used to analyze the coupling constants. It can be divided into two parts,
\begin{eqnarray}
\Pi^{\mathrm{OPE}}=\Pi^{\mathrm{pert}}+\Pi^{\mathrm{non-pert}}
\end{eqnarray}
where $\Pi^{\mathrm{pert}}$ refers to the perturbative part and $\Pi^{\mathrm{non-pert}}$ denotes the non-perturbative contributions including $\langle\bar{q}q\rangle$, $\langle g_{s}^{2}G^{2}\rangle$, $\langle\bar{q}g_{s}\sigma Gq\rangle$, $\langle f^{3}G^{3}\rangle$ and $\langle\bar{q}q\rangle\langle g_{s}^{2}G^{2}\rangle$.  The perturbative part, $\langle g_{s}^{2}G^{2}\rangle$ and $\langle f^{3}G^{3}\rangle$ condensate terms can be written as the following form by using the dispersion relation,
\begin{eqnarray}
\notag
\Pi }(p,p') =  - \int\limits_0^\infty  {\int\limits_0^\infty  {\frac{{{\rho }(s,u,{q^2})}}{{(s - {p^2})(u - p{'^2})}}dsdu}
\end{eqnarray}
where $\rho(s,u,q^{2})$ is the QCD spectral density,
\begin{eqnarray}
\notag
\rho (s,u,{q^2}) &&= {\rho ^{\mathrm{pert}}}(s,u,{q^2}) + {\rho ^{\left\langle {g_s^2{G^2}} \right\rangle }}(s,u,{q^2}) \\
&&+ {\rho ^{\left\langle {{f^3}{G^3}} \right\rangle }}(s,u,{q^2})
\end{eqnarray}
and $s=p^{2}$, $u=p'^{2}$, and $q=p-p'$.

For vertex $DDJ/\psi$ as an example, we show how the QCD spectral density is obtained. For the perturbative contribution, we firstly substitute the free propagator in the momentum space in Eq. (\ref{eq:7}). After performing integrations in the coordinate and momentum space, the correlation functions are expressed as,
\begin{eqnarray}
\notag
&&\Pi _\mu ^{J/\psi }(p,p') = \frac{{3{i^3}}}{{{{(2\pi )}^4}}}\int {{d^4}k} \\
\notag
&& \times \frac{{Tr[(\cancel k + \cancel q + {m_Q}){\gamma _5}(\cancel k - \cancel p' + {m_q}){\gamma _5}(\cancel k + {m_Q}){\gamma _\mu }]}}{{[{{(k + q)}^2} - m_Q^2][{{(k - p')}^2} - m_q^2]({k^2} - m_Q^2)}}\\
\notag
&&\Pi _\mu ^D(p,p') = \frac{{3{i^3}}}{{{{(2\pi )}^4}}}\int {{d^4}k} \\
&& \times \frac{{Tr[(\cancel k - \cancel p' - {m_Q}){\gamma _5}(\cancel k + \cancel q){\gamma _5}(\cancel k - {m_Q}){\gamma _\mu }]}}{{[{{(k - p')}^2} - m_Q^2][{{(k + q)}^2} - m_q^2]({k^2} + m_Q^2)}}
\end{eqnarray}

\begin{figure}[htbp]
\centering
\includegraphics[width=9cm]{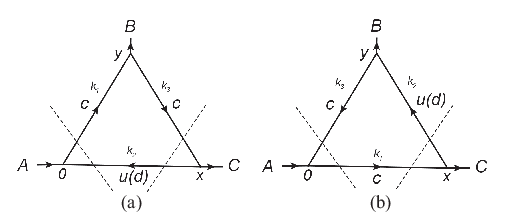}
\caption{The perturbative contributions for $J/\psi(\eta_{c})$ (a) and $D(D^{*})$ (b) off-shell. The dashed lines denote the Cutkosky cuts.}
\label{fig:free}
\end{figure}

Then, we put all the quark lines on mass-shell using the Cutkosky's rules(Fig. \ref{fig:free}), and the QCD spectral density for the perturbative contribution will be obtained,
\begin{eqnarray}
\notag
&&\rho _\mu ^{J/\psi }(s,u,{q^2}) = \frac{{3{i^3}}}{{{{(2\pi )}^4}}}\int {{d^4}k} \delta [{(k + q)^2} - m_Q^2]\\
\notag
&& \times \delta [{(k - p')^2} - m_q^2]\delta ({k^2} - m_Q^2)Tr[(\cancel k + \cancel q + {m_Q}){\gamma _5}\\
\notag
&& \times (\cancel k - \cancel p' + {m_q}){\gamma _5}(\cancel k + {m_Q}){\gamma _\mu }]\\
\notag
&& =  - \frac{3}{{{{(2\pi )}^3}}}\frac{\pi }{{2\sqrt {\lambda (s,u,{q^2})} }}Tr\{ [({C_p} + 1)\cancel p + ({C_{p'}} - 1)\cancel p'\\
\notag
&& + {m_Q}]{\gamma _5}[{C_p}\cancel p + ({C_{p'}} - 1)\cancel p' + {m_q}]{\gamma _5}\\
\notag
&& \times ({C_p}\cancel p + {C_{p'}}\cancel p' + {m_Q}){\gamma _\mu }\}
\end{eqnarray}
\begin{eqnarray}\label{eq:des}
\notag
&&\rho _\mu ^D(s,u,q^2) = \frac{{3{i^3}}}{{{{(2\pi )}^4}}}\int {{d^4}k} \delta [{(k - p')^2} - m_Q^2]\\
\notag
&& \times \delta [{(k + q)^2} - m_q^2]\delta ({k^2} + m_Q^2)Tr[(\cancel k - \cancel p' - {m_Q}){\gamma _5}\\
\notag
&& \times (\cancel k + \cancel q - {m_q}){\gamma _5}(\cancel k - {m_Q}){\gamma _\mu }]\\
\notag
&& = \frac{3}{{{{(2\pi )}^3}}}\frac{\pi }{{2\sqrt {\lambda (s,u,{q^2})} }}Tr\{ [{C'_p}\cancel p + ({C'_{p'}} - 1)\cancel p'\\
\notag
&& - {m_Q}]{\gamma _5}[({C'_p} + 1)\cancel p + ({C'_{p'}} - 1)\cancel p' - {m_q}]{\gamma _5}\\
&& \times ({C'_p}\cancel p + {C'_{p'}}\cancel p' - {m_Q}){\gamma _\mu }\}
\end{eqnarray}
where,
\begin{eqnarray}
\notag
&&{C_p} = \frac{{(u + m_Q^2)(s + u - {q^2}) - 2u(u - {q^2} + m_Q^2)}}{{\lambda (s,u,{q^2})}}\\
\notag
&&{C_{p'}} = \frac{{(u - {q^2} + m_Q^2)(s + u - {q^2}) - 2s(u + m_Q^2)}}{{\lambda (s,u,{q^2})}}\\
\notag
&&{C'_p} = \frac{{u(s + u - {q^2}) - 2u(u - {q^2} - m_Q^2)}}{{\lambda (s,u,{q^2})}}\\
\notag
&&{C'_{p'}} = \frac{{(u - {q^2} - m_Q^2)(s + u - {q^2}) - 2su}}{{\lambda (s,u,{q^2})}}\\
&&\lambda (s,u,{q^2}) = {(s + u - {q^2})^2} - 4su
\end{eqnarray}
As for the vacuum condensate terms $\langle g_{s}^{2}G^{2}\rangle$ and $\langle f^{3}G^{3}\rangle$, a typical integral will be encountered,
\begin{eqnarray}
I_{ijk} = \int {{d^4}k\frac{1}{{{{[{{(k + q)}^2} - m_1^2]}^i}{{[{{(k - p')}^2} - m_2^2]}^j}{{({k^2} - m_3^2)}^k}}}}
\end{eqnarray}
According to the following transformations, these condensate terms can also be calculated by Cutkosky's rules,
\begin{eqnarray}
\notag
&&I_{ijk} = \frac{1}{{(i - 1)!(j - 1)!(k - 1)!}}\frac{{{\partial ^{i - 1}}}}{{\partial {A^{i - 1}}}}\frac{{{\partial ^{j - 1}}}}{{\partial {B^{j - 1}}}}\frac{{{\partial ^{k - 1}}}}{{\partial {C^{k - 1}}}}\int {{d^4}k} \\
\notag
&& \times \frac{1}{{[{{(k + q)}^2} - A][{{(k - p')}^2} - B]({k^2} - C)}}{|_{A \to {m_{1,}}B \to {m_{2,}}C \to {m_3}}}\\
\notag
&& \to \frac{{{{( - 2\pi i)}^3}}}{{{{(2\pi i)}^2}}}\frac{1}{{(i - 1)!(j - 1)!(k - 1)!}}\frac{{{\partial ^{i - 1}}}}{{\partial {A^{i - 1}}}}\frac{{{\partial ^{j - 1}}}}{{\partial {B^{j - 1}}}}\frac{{{\partial ^{k - 1}}}}{{\partial {C^{k - 1}}}} \\
\notag
&& \times \int {{d^4}k} \delta [{(k + q)^2} - A]\delta [{(k - p')^2} - B]\\
\notag
&&\times \delta ({k^2} - C){|_{A \to {m_{1,}}B \to {m_{2,}}C \to {m_3}}} \\
\notag
&& = \frac{{{{( - 2\pi i)}^3}}}{{{{(2\pi i)}^2}}}\frac{1}{{(i - 1)!(j - 1)!(k - 1)!}}\frac{{{\partial ^{i - 1}}}}{{\partial {A^{i - 1}}}}\frac{{{\partial ^{j - 1}}}}{{\partial {B^{j - 1}}}}\frac{{{\partial ^{k - 1}}}}{{\partial {C^{k - 1}}}}\\
&& \times \frac{\pi }{{2\sqrt {\lambda (s,u,{q^2})} }}{|_{A \to {m_{1,}}B \to {m_{2,}}C \to {m_3}}}
\end{eqnarray}

\begin{figure*}[htbp]
\centering
\subfigure[]{\includegraphics[height=3.2cm,width=3.5cm]{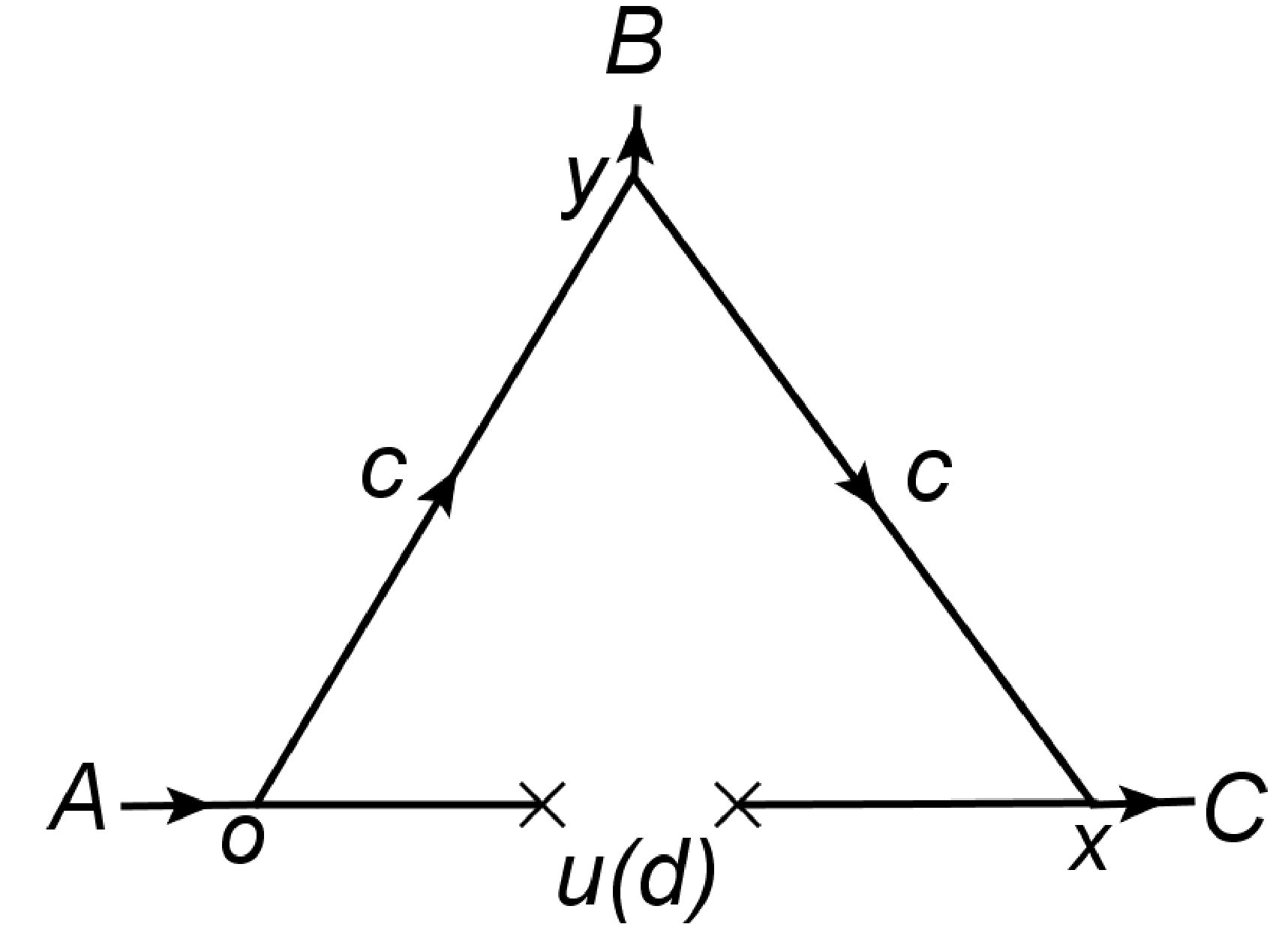}}
\subfigure[]{\includegraphics[height=3.2cm,width=3.5cm]{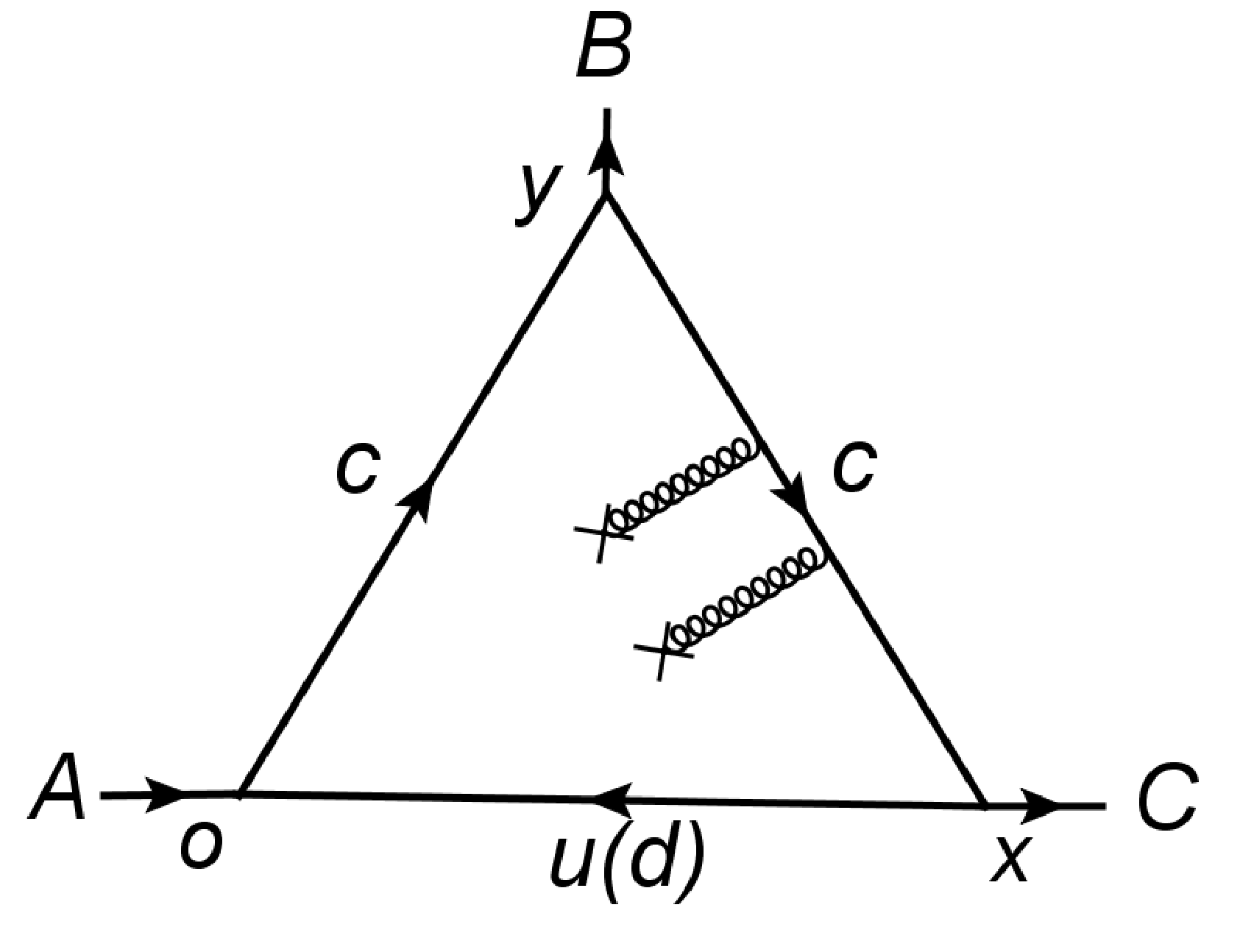}}
\subfigure[]{\includegraphics[height=3.2cm,width=3.5cm]{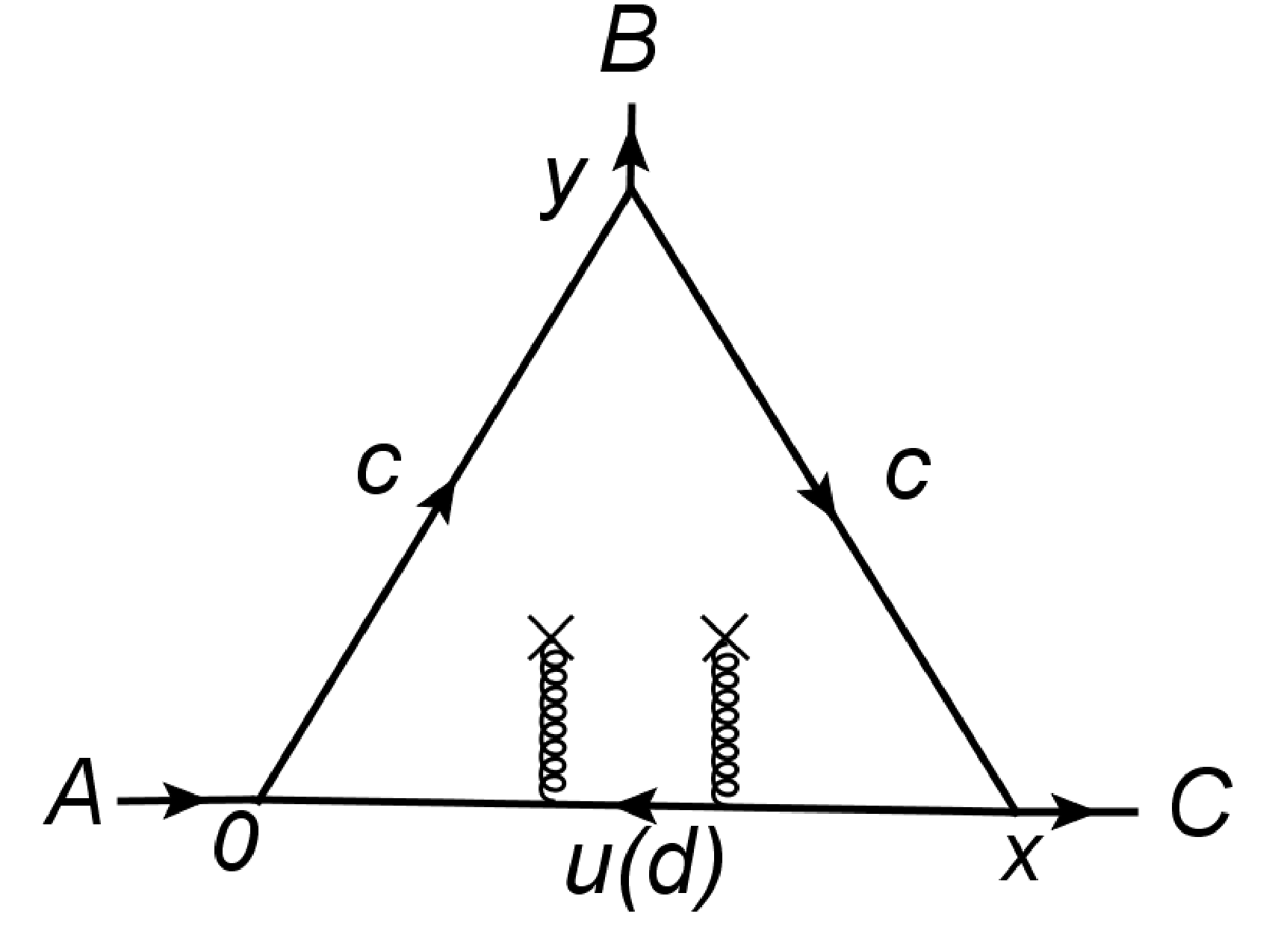}}
\subfigure[]{\includegraphics[height=3.2cm,width=3.5cm]{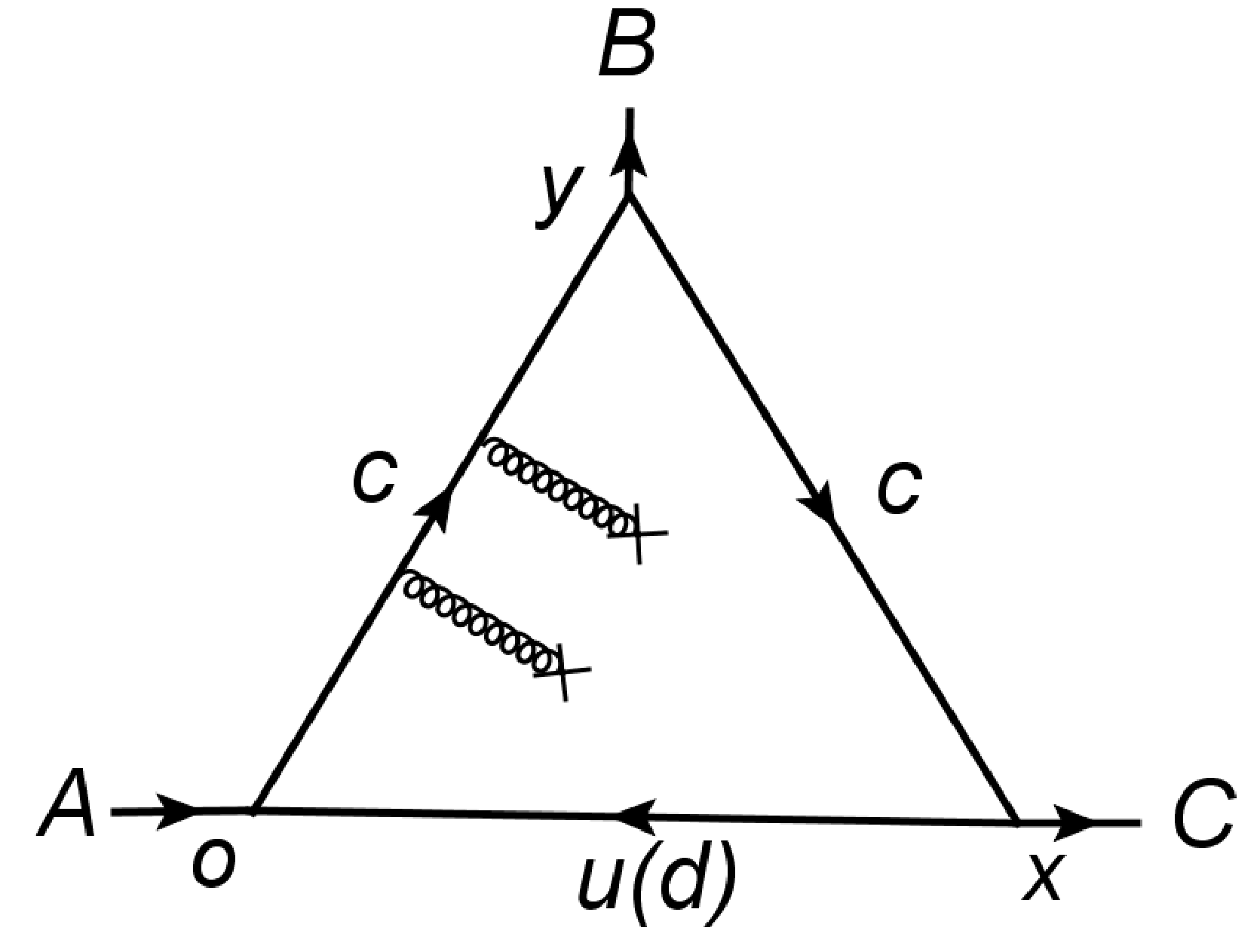}}
\subfigure[]{\includegraphics[height=3.2cm,width=3.5cm]{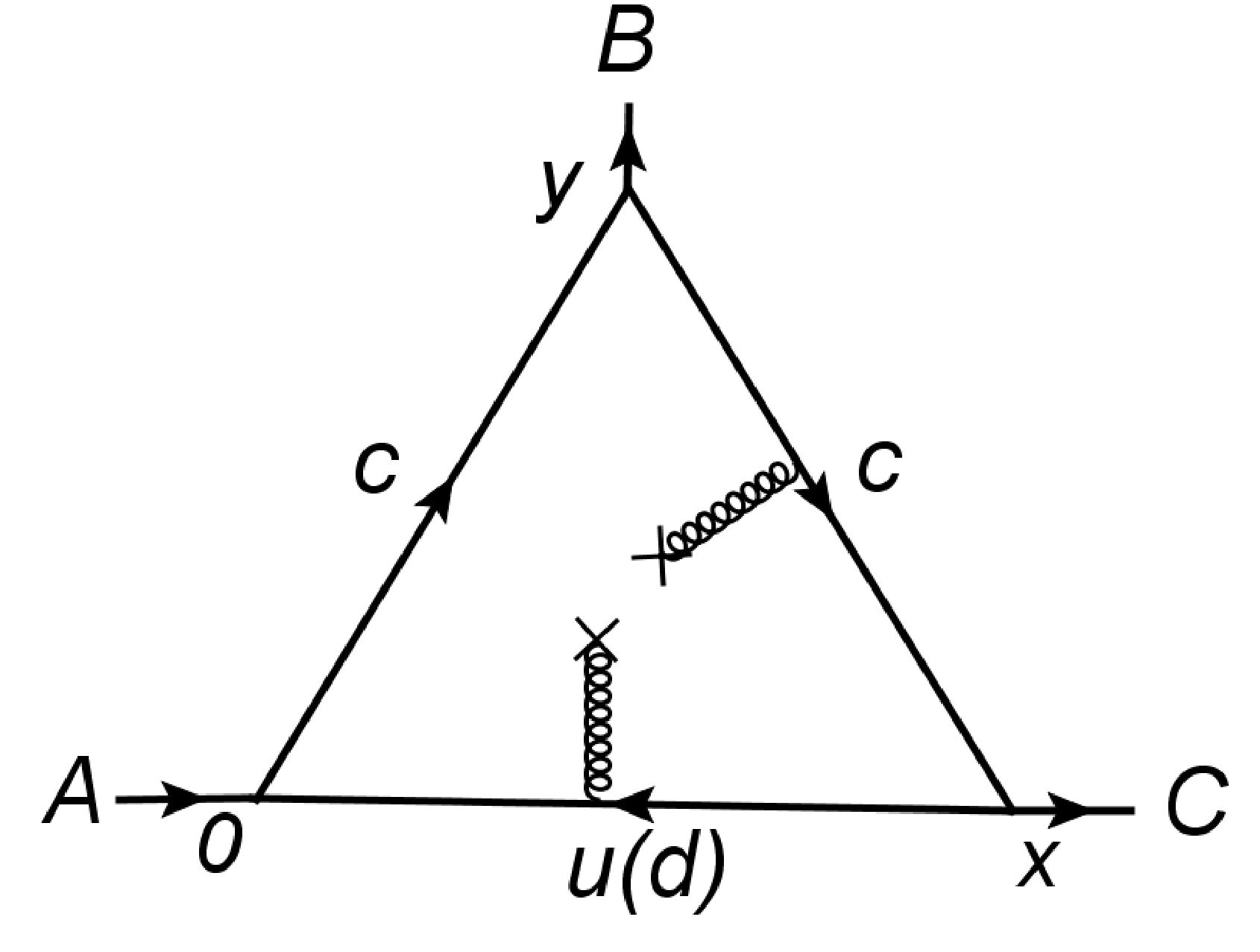}}

\subfigure[]{\includegraphics[height=3.2cm,width=3.5cm]{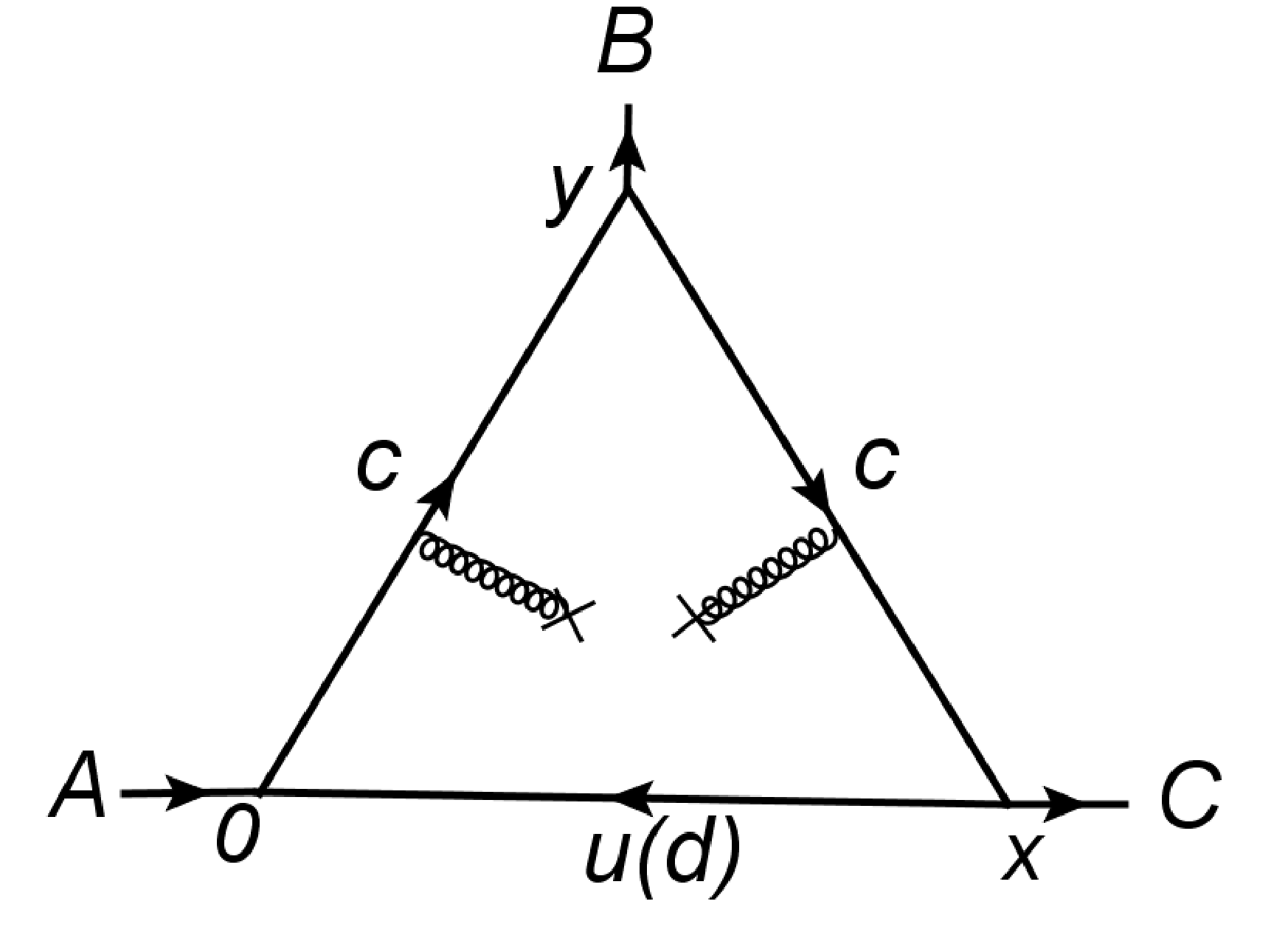}}
\subfigure[]{\includegraphics[height=3.2cm,width=3.5cm]{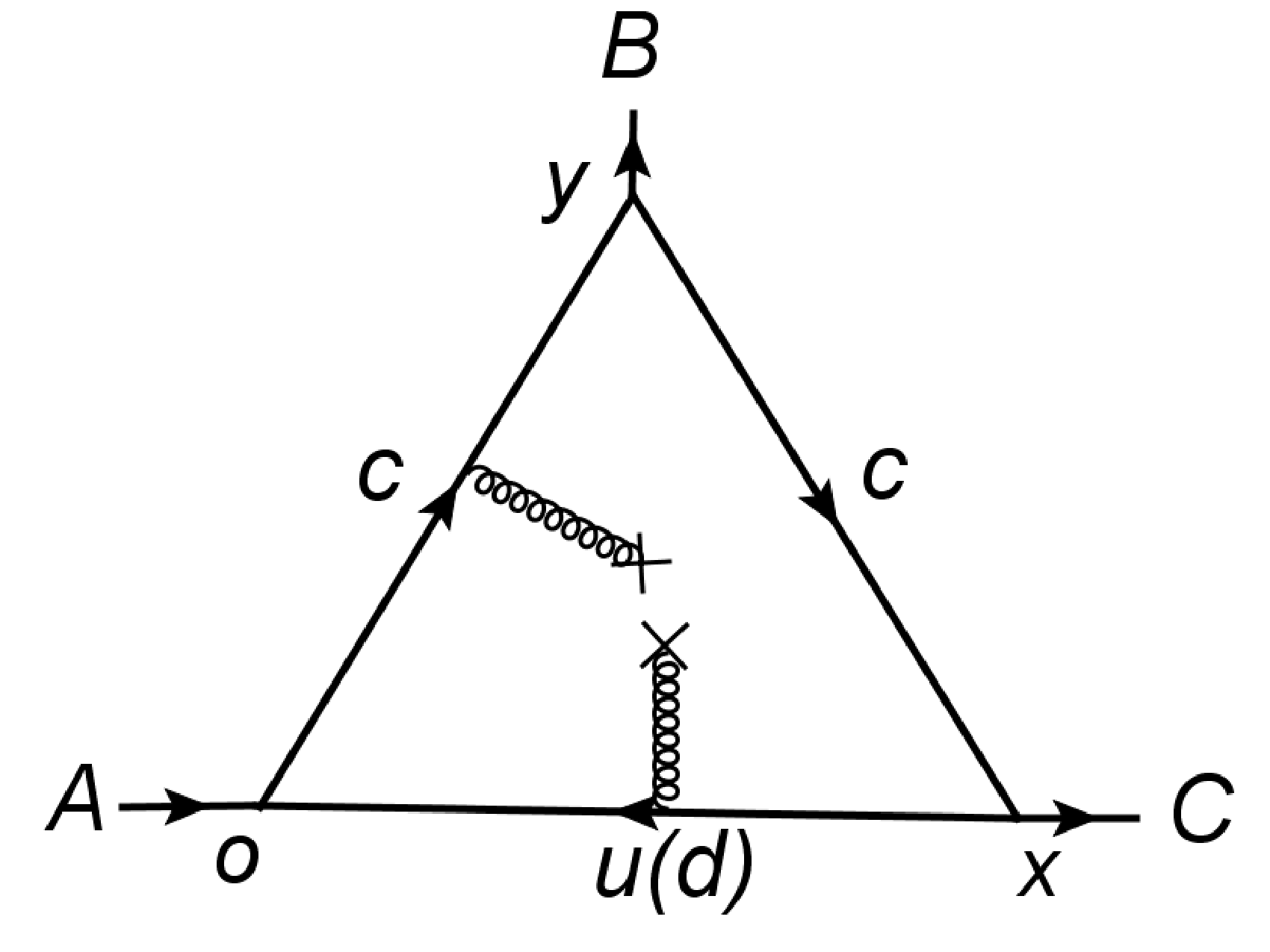}}
\subfigure[]{\includegraphics[height=3.2cm,width=3.5cm]{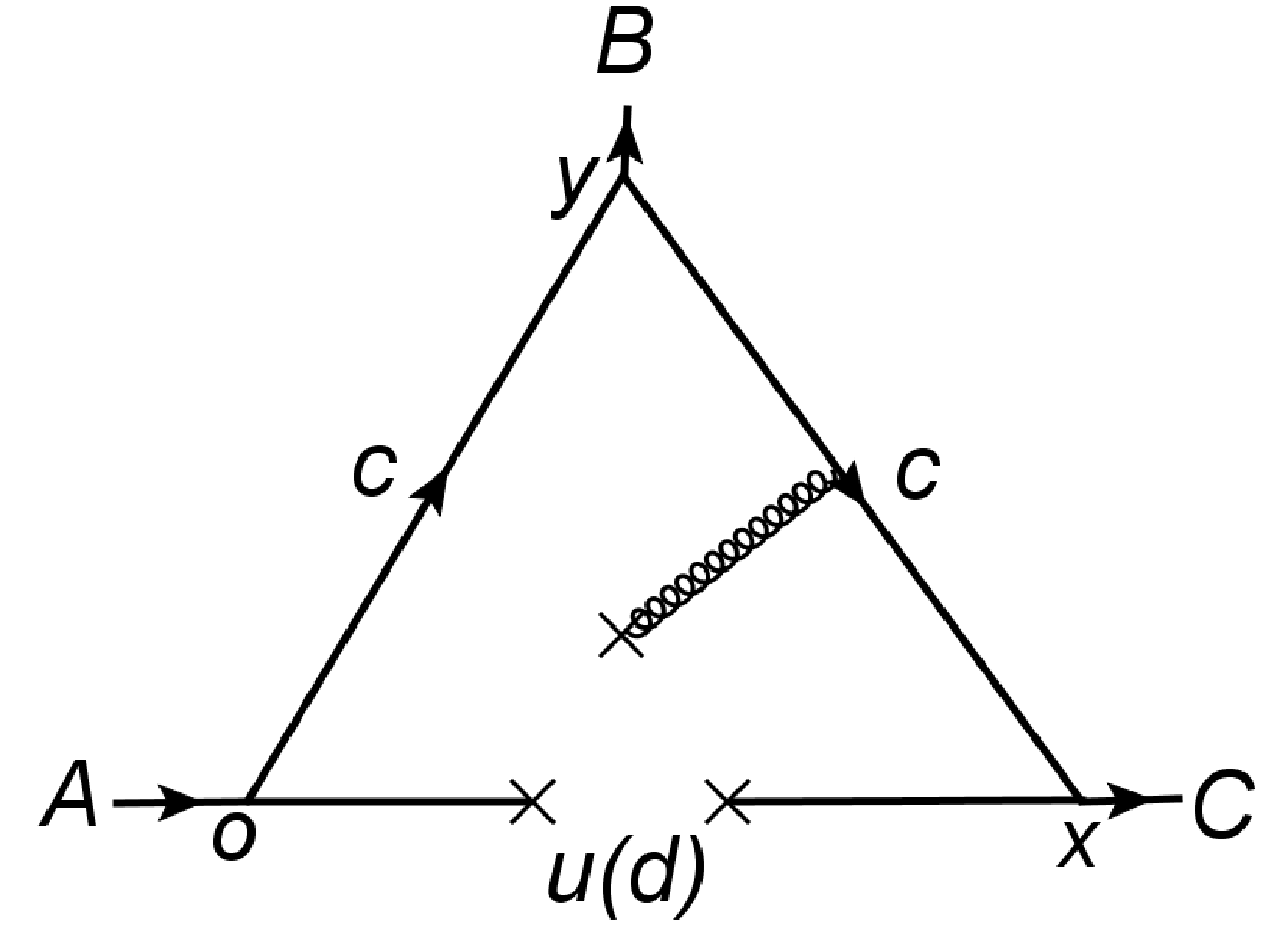}}
\subfigure[]{\includegraphics[height=3.2cm,width=3.5cm]{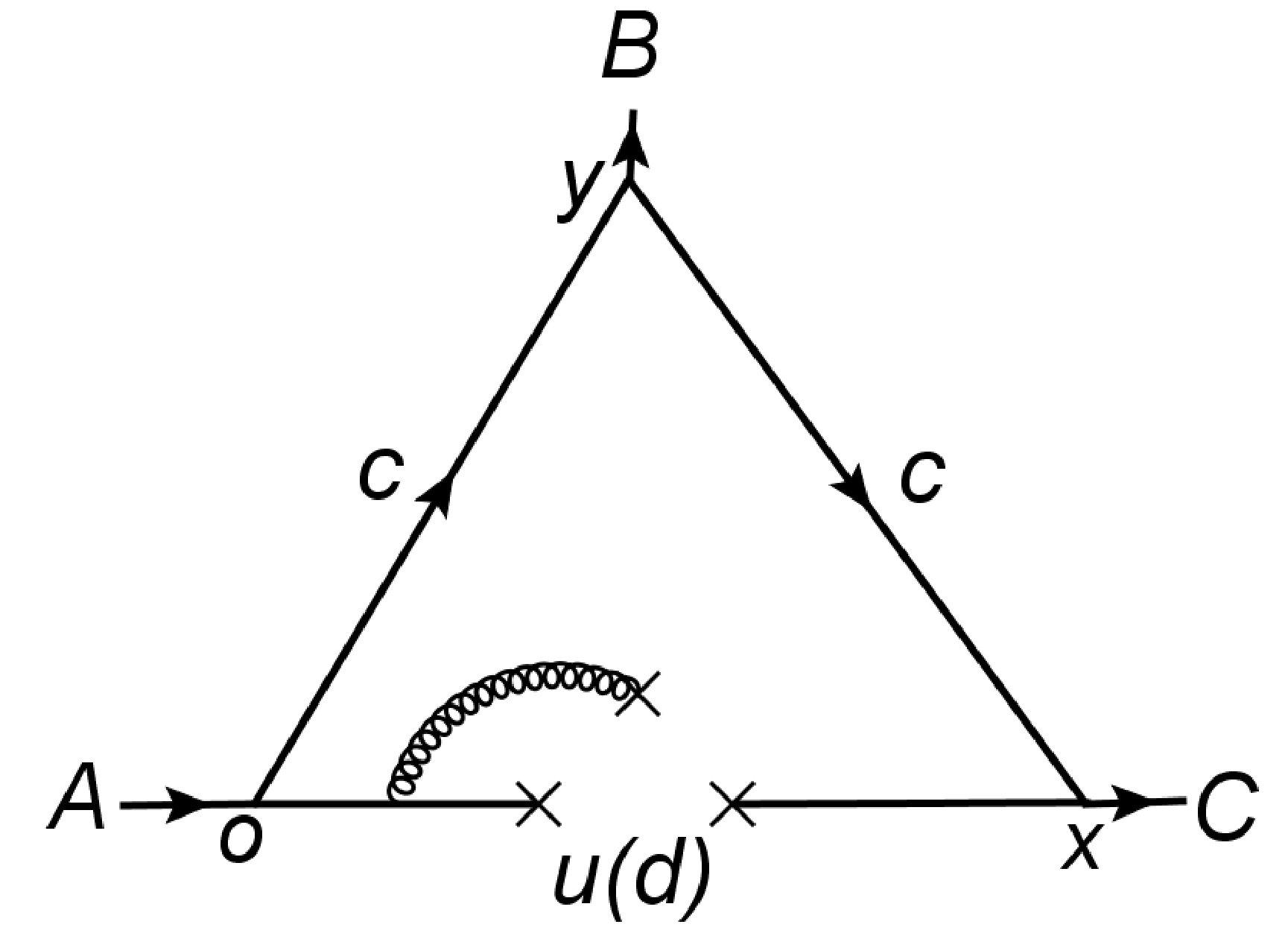}}
\subfigure[]{\includegraphics[height=3.2cm,width=3.5cm]{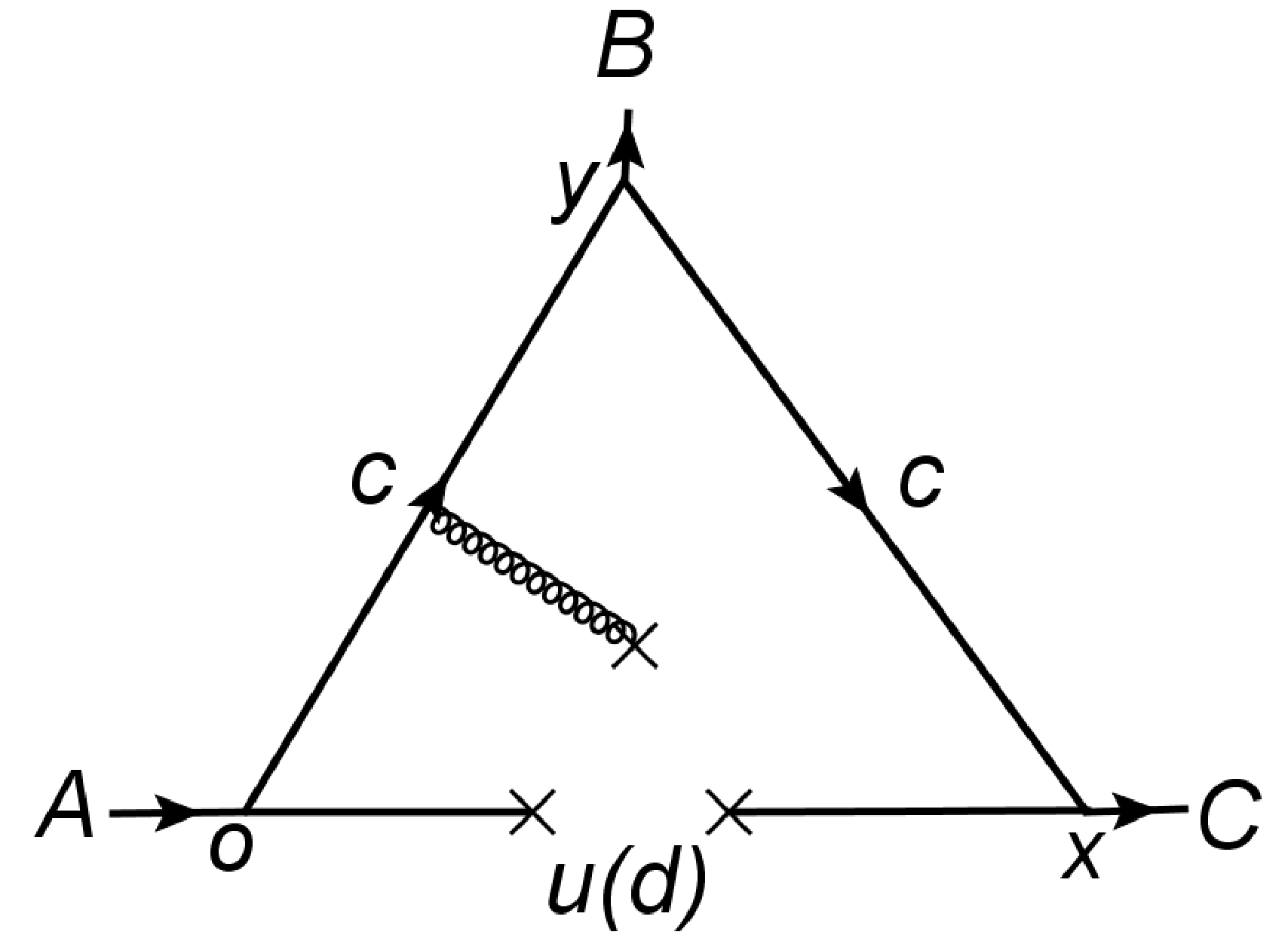}}

\subfigure[]{\includegraphics[height=3.2cm,width=3.5cm]{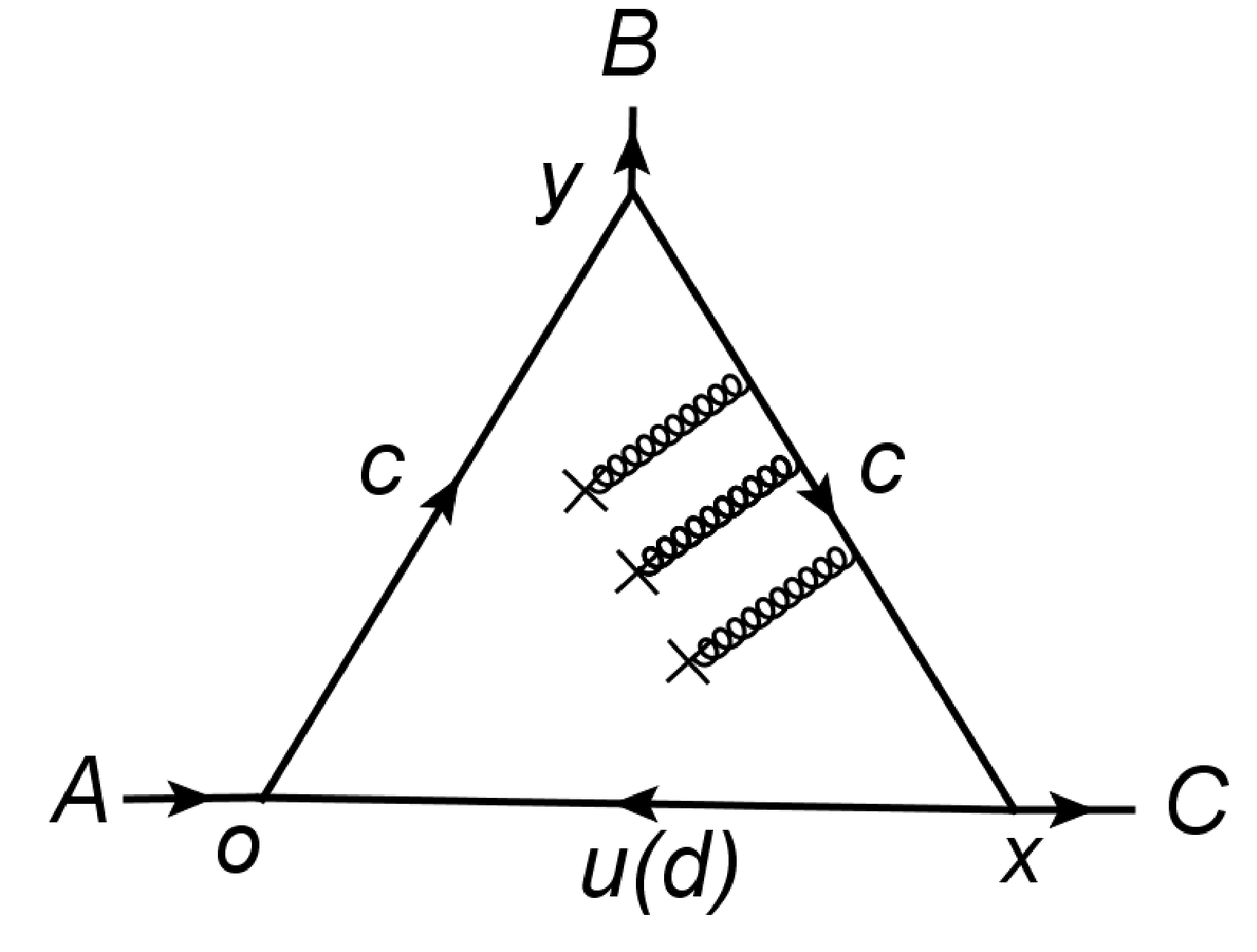}}
\subfigure[]{\includegraphics[height=3.2cm,width=3.5cm]{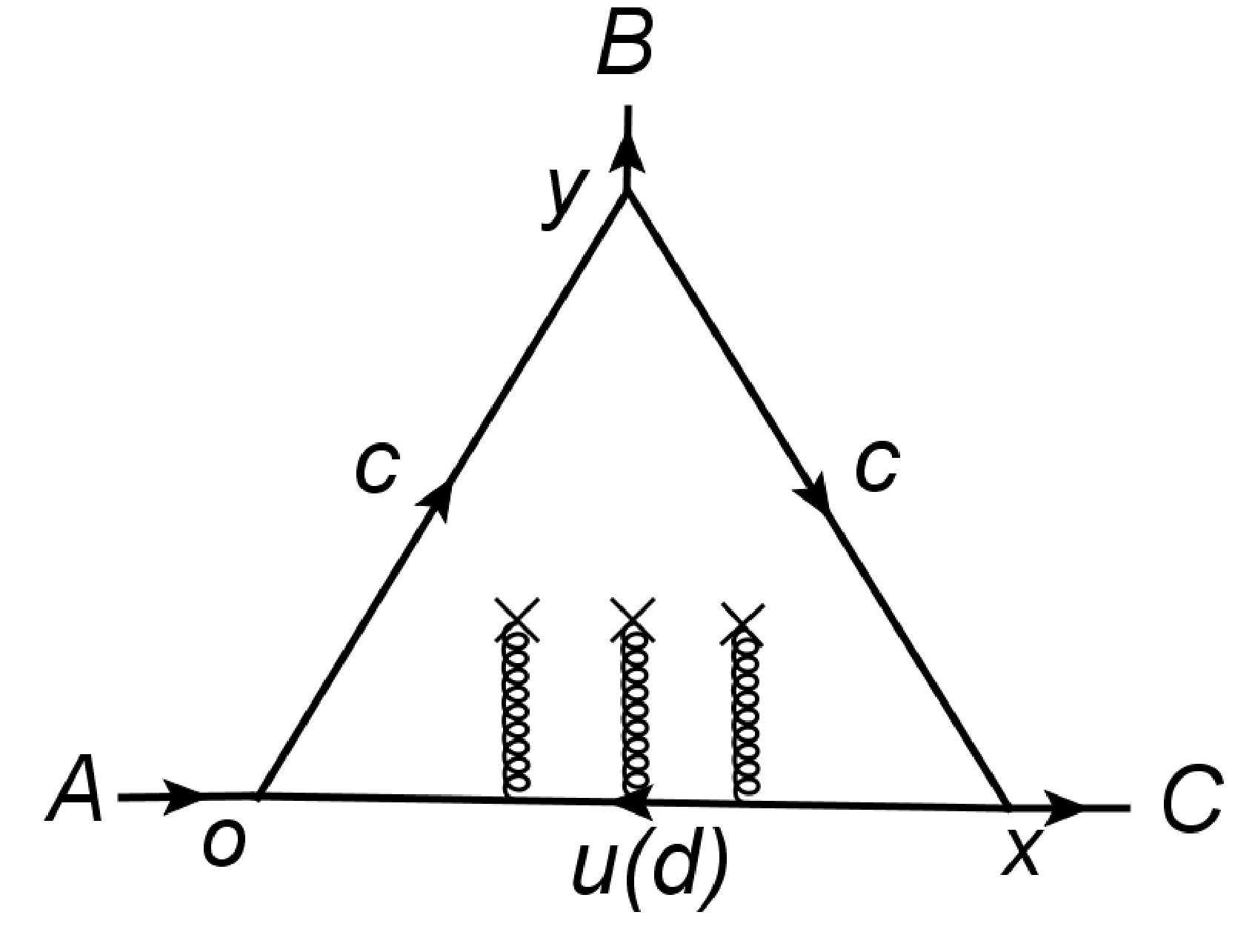}}
\subfigure[]{\includegraphics[height=3.2cm,width=3.5cm]{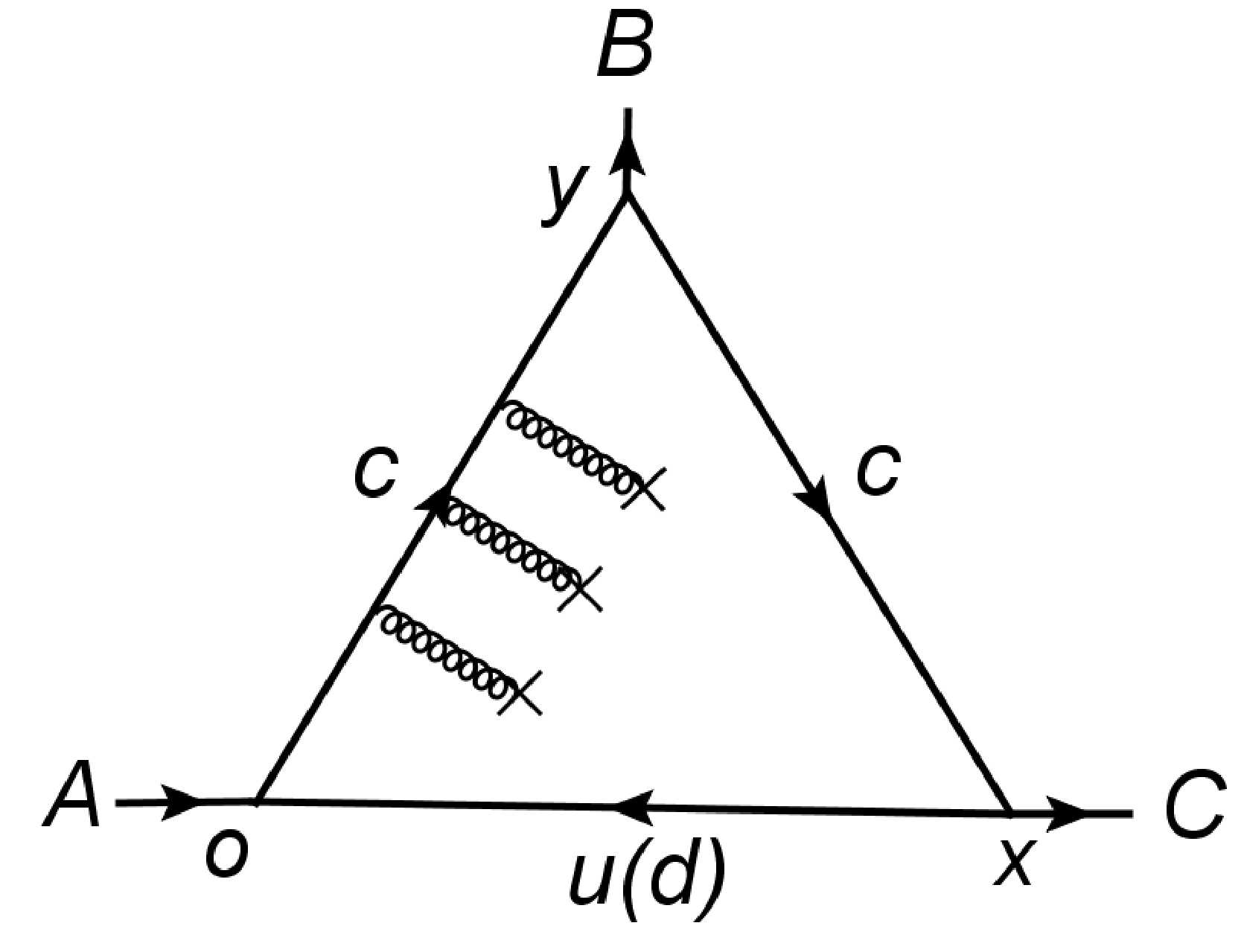}}
\subfigure[]{\includegraphics[height=3.2cm,width=3.5cm]{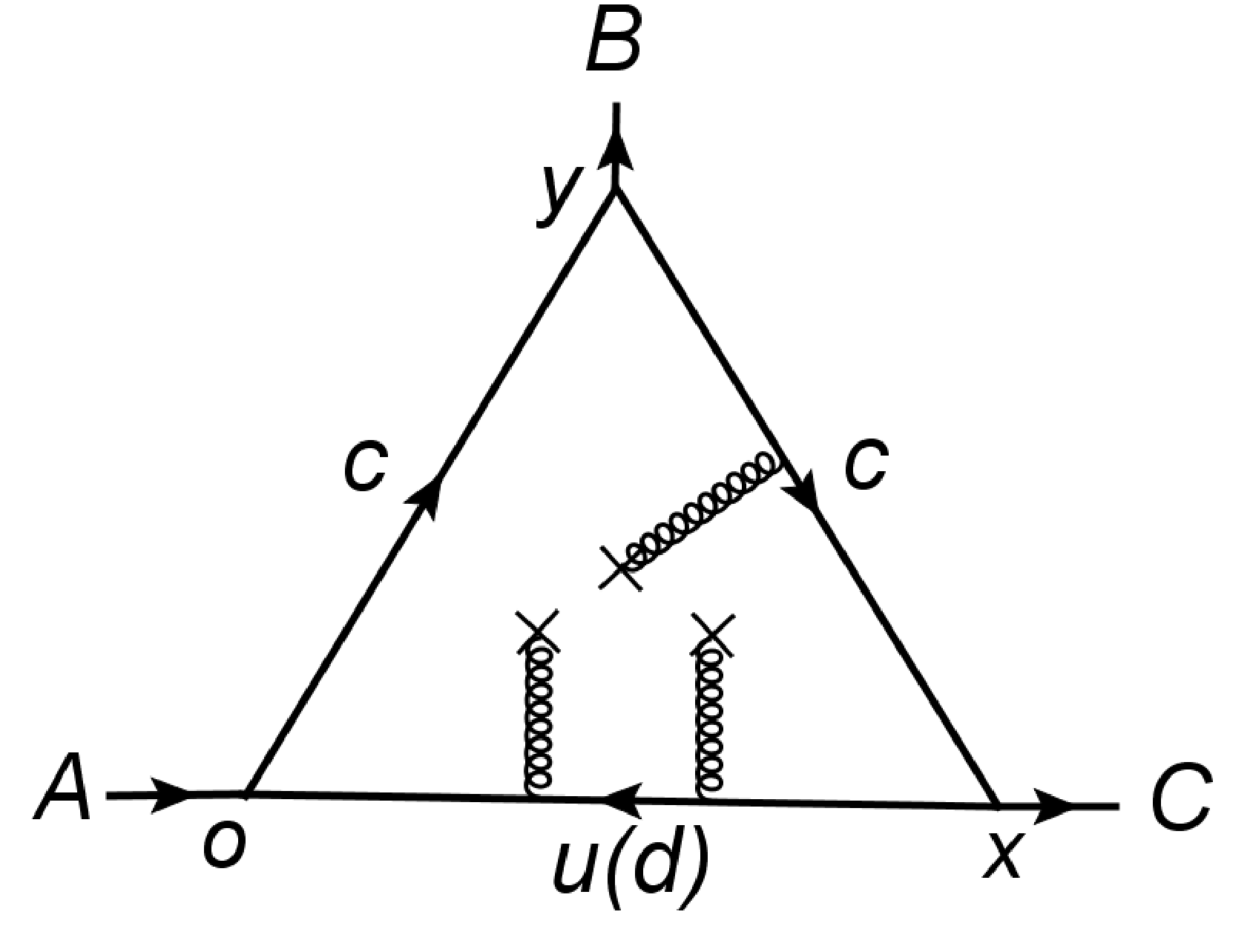}}
\subfigure[]{\includegraphics[height=3.2cm,width=3.5cm]{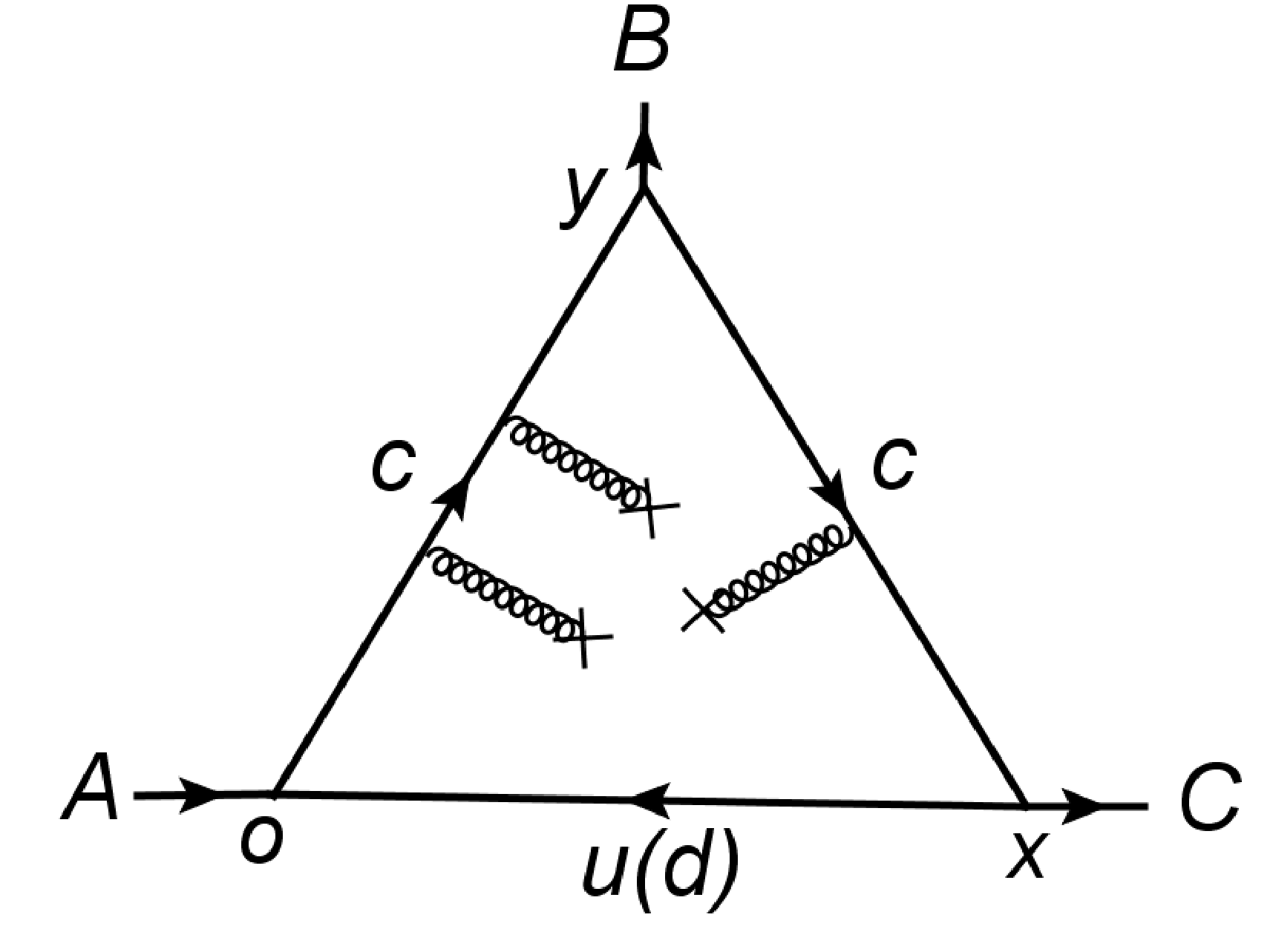}}

\subfigure[]{\includegraphics[height=3.2cm,width=3.5cm]{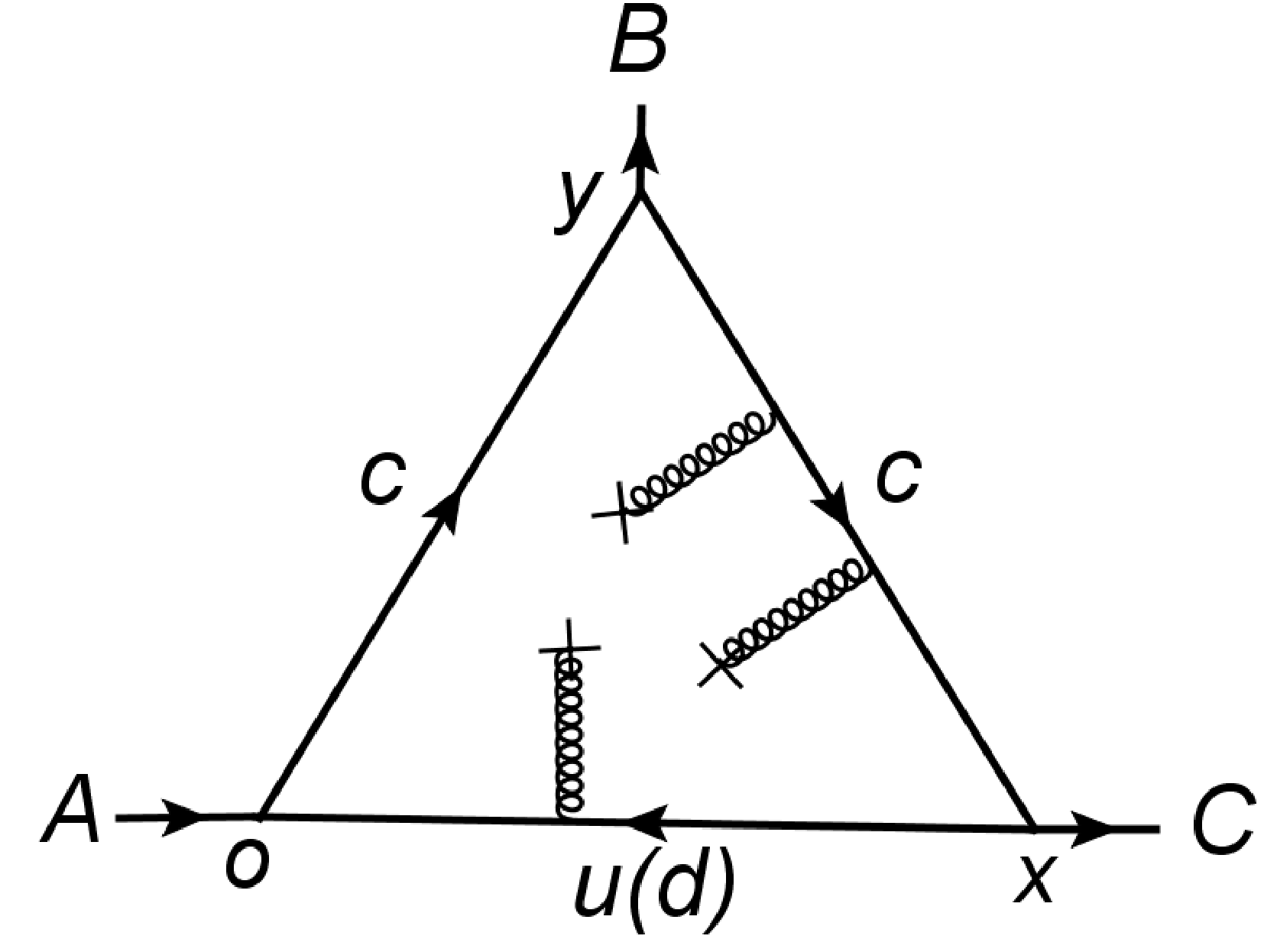}}
\subfigure[]{\includegraphics[height=3.2cm,width=3.5cm]{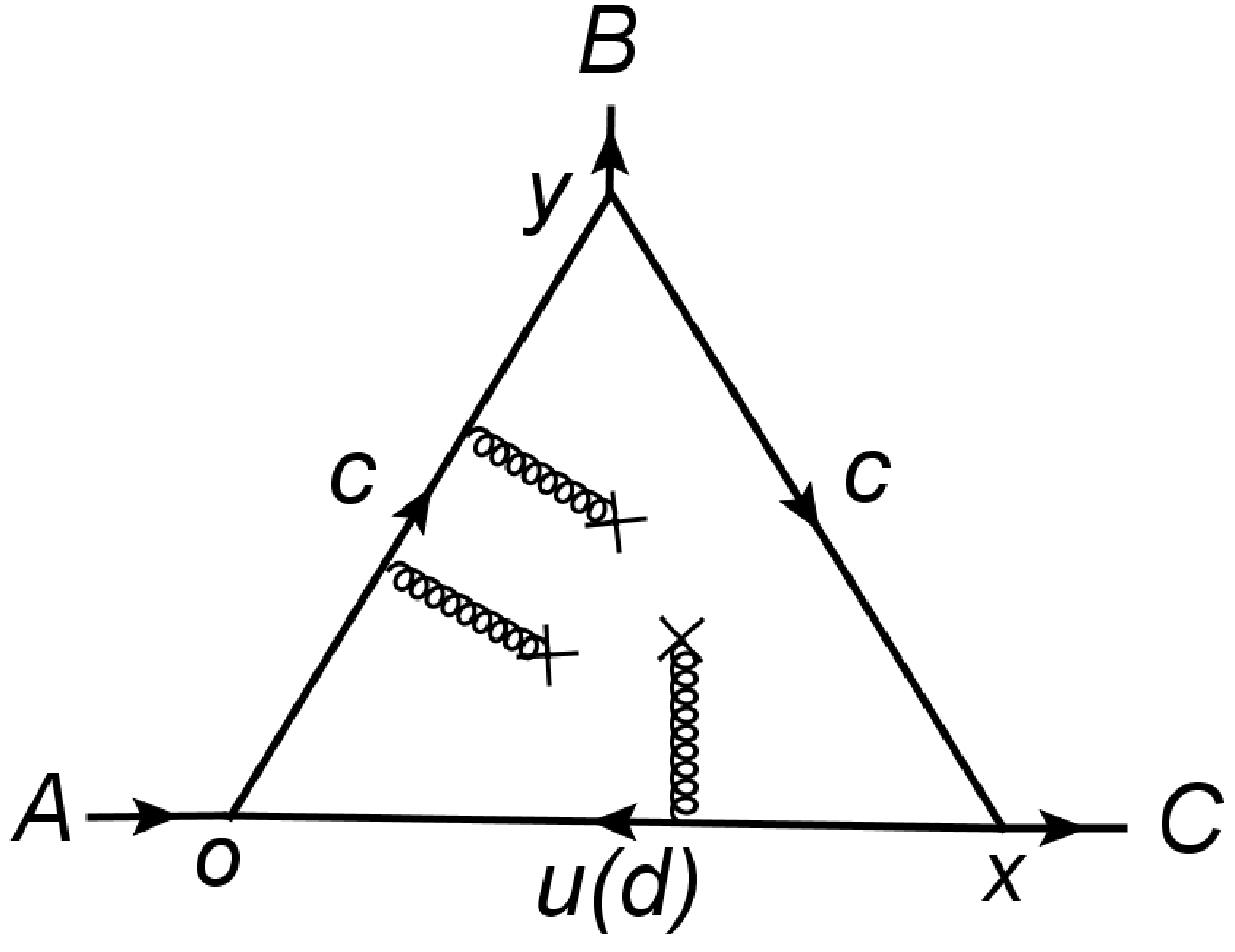}}
\subfigure[]{\includegraphics[height=3.2cm,width=3.5cm]{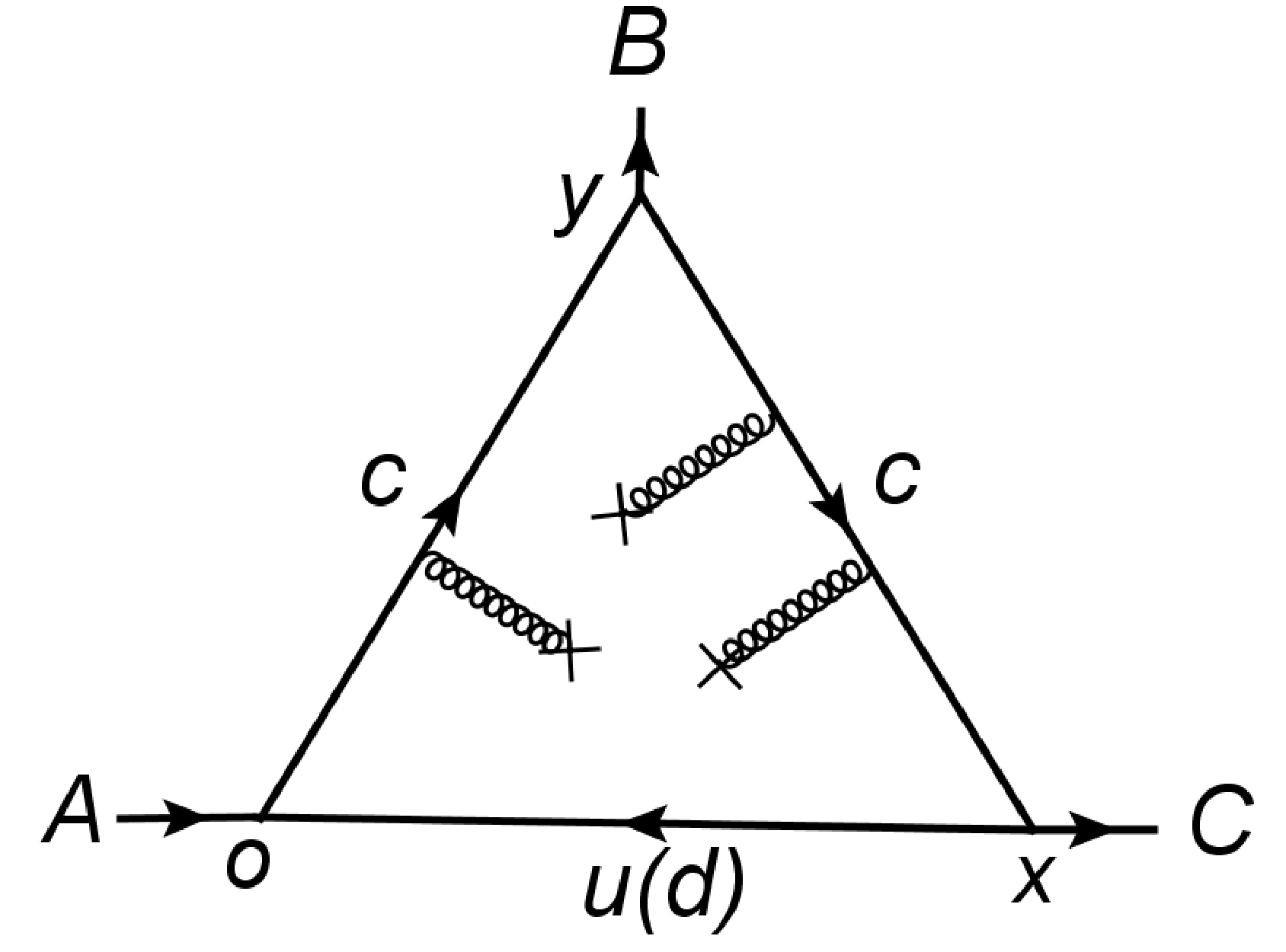}}
\subfigure[]{\includegraphics[height=3.2cm,width=3.5cm]{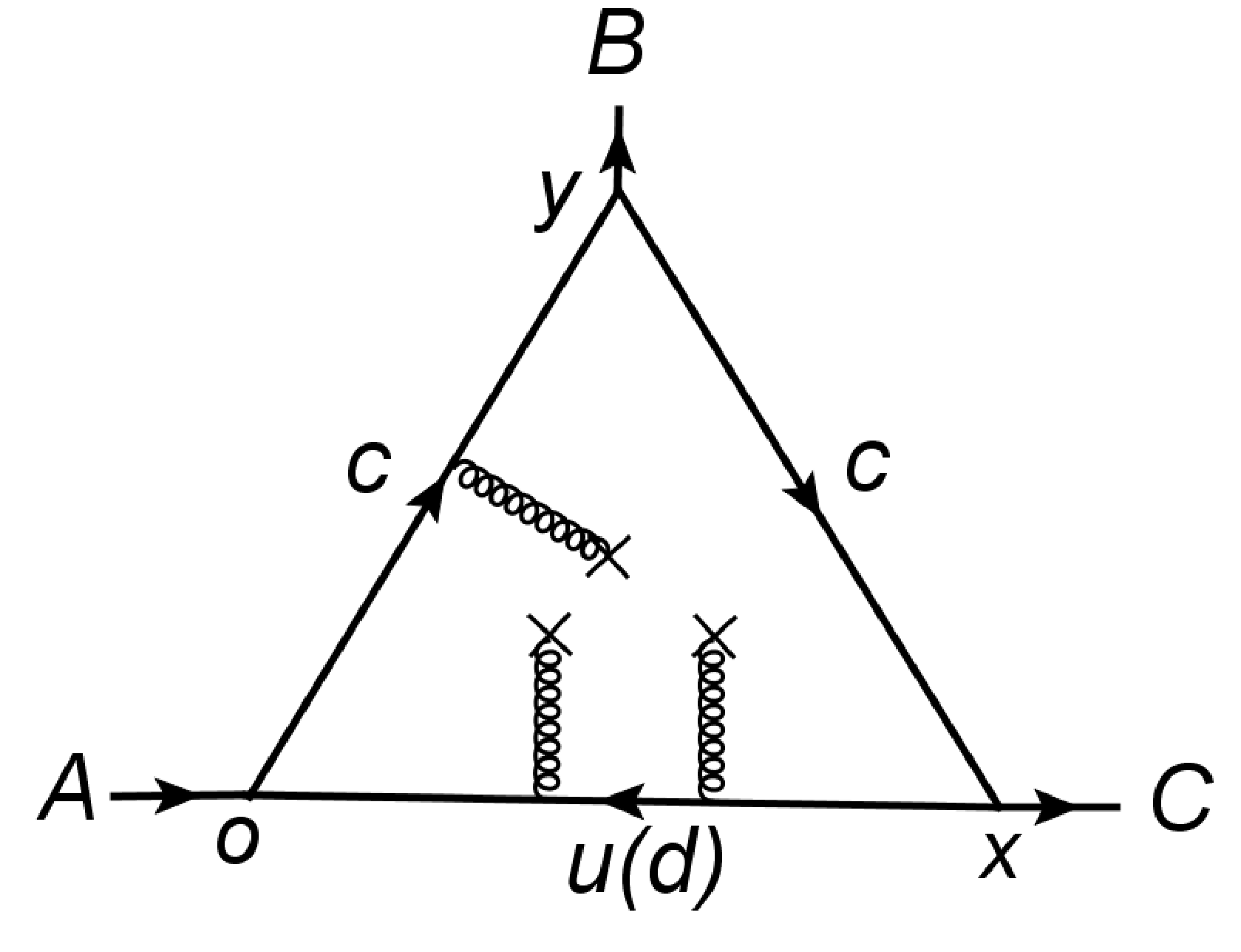}}
\subfigure[]{\includegraphics[height=3.2cm,width=3.5cm]{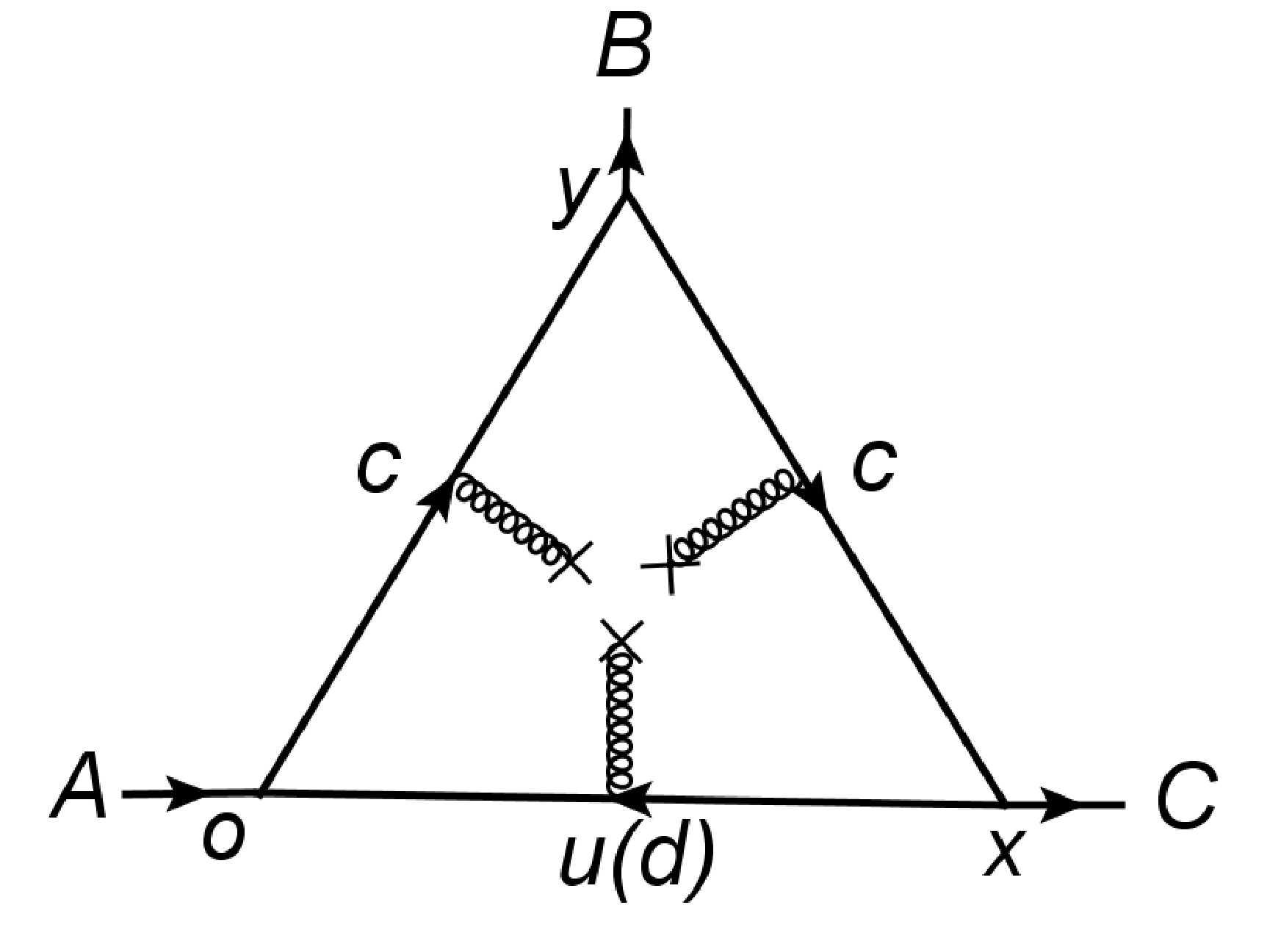}}

\subfigure[]{\includegraphics[height=3.2cm,width=3.5cm]{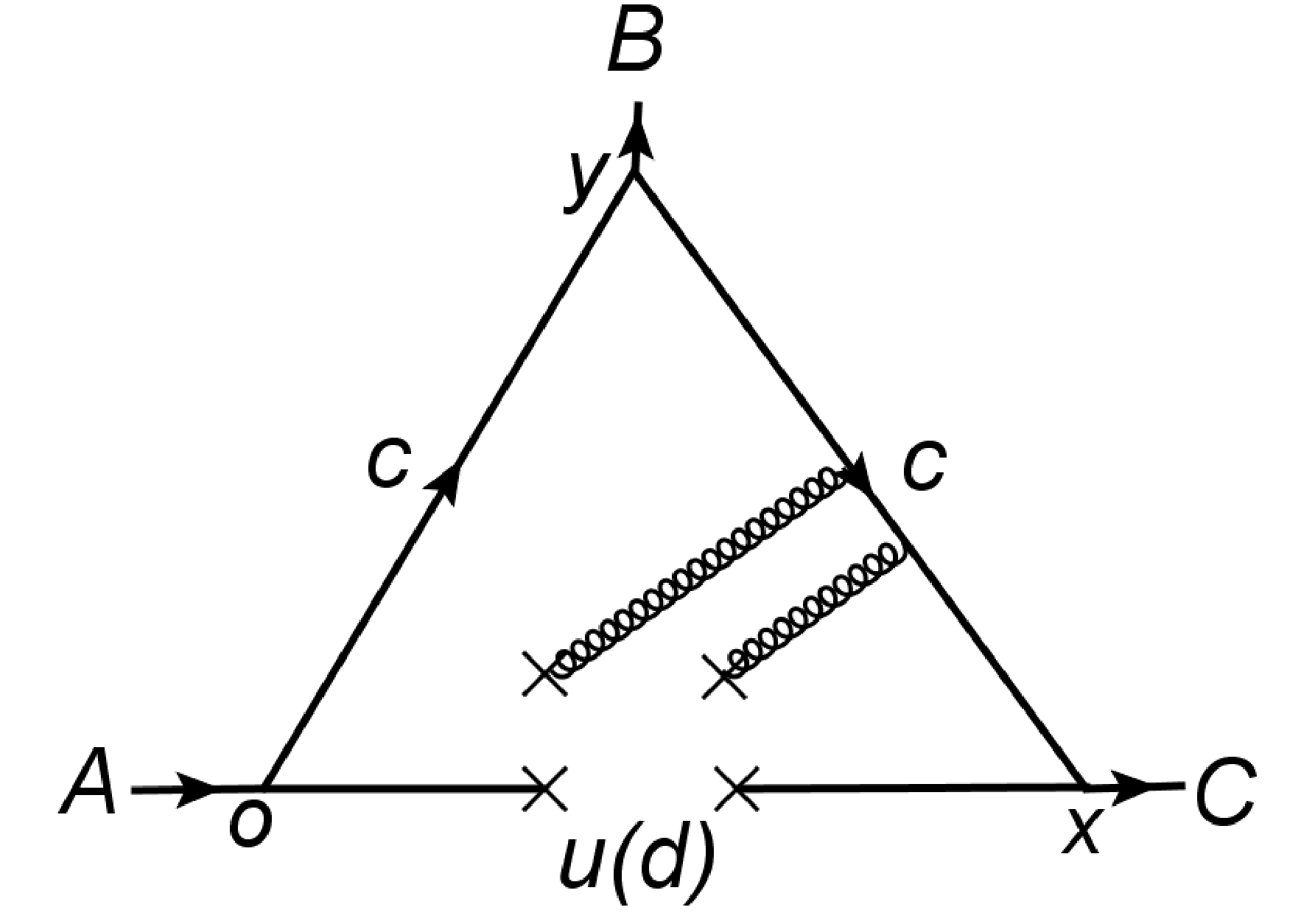}}
\subfigure[]{\includegraphics[height=3.2cm,width=3.5cm]{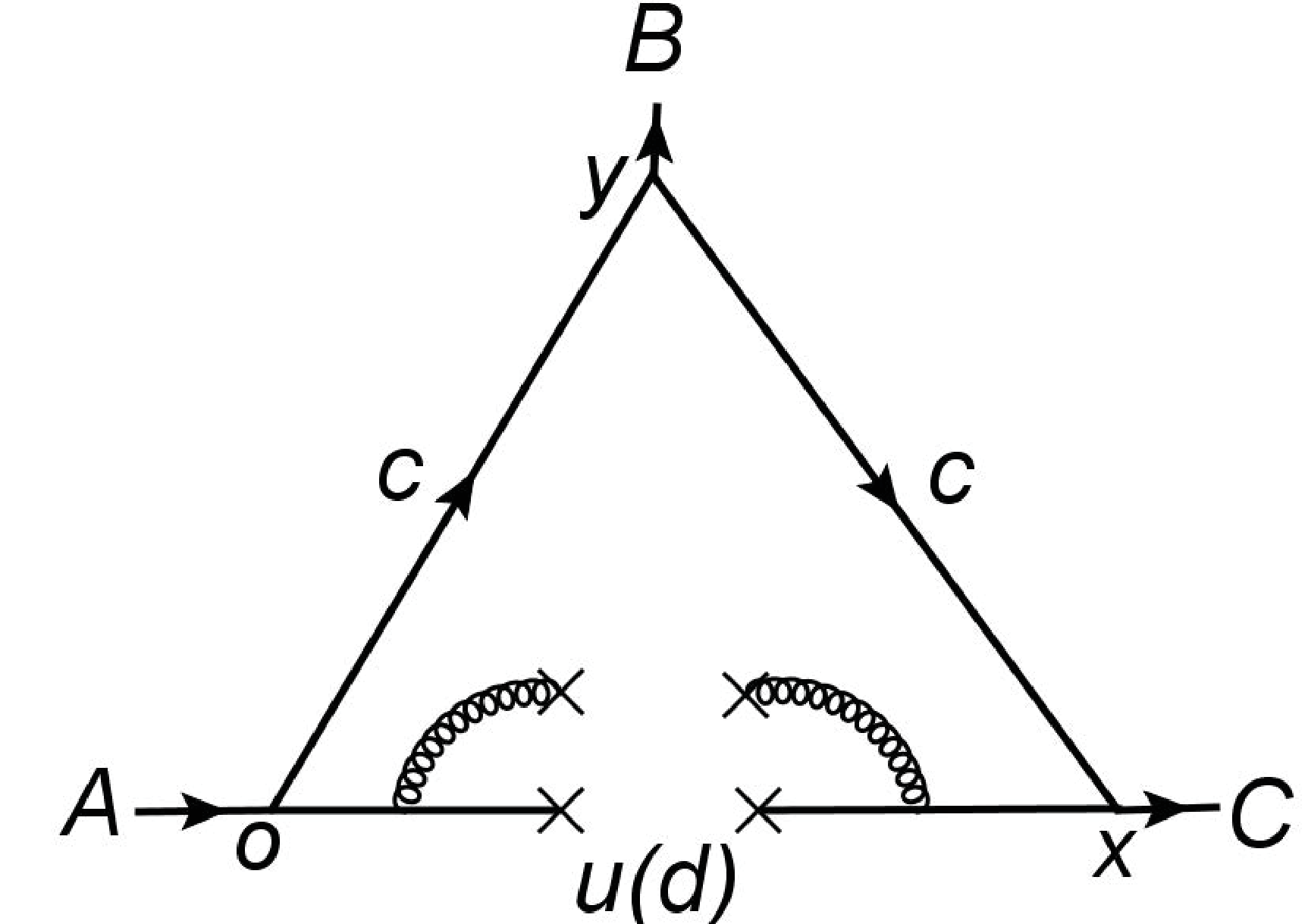}}
\subfigure[]{\includegraphics[height=3.2cm,width=3.5cm]{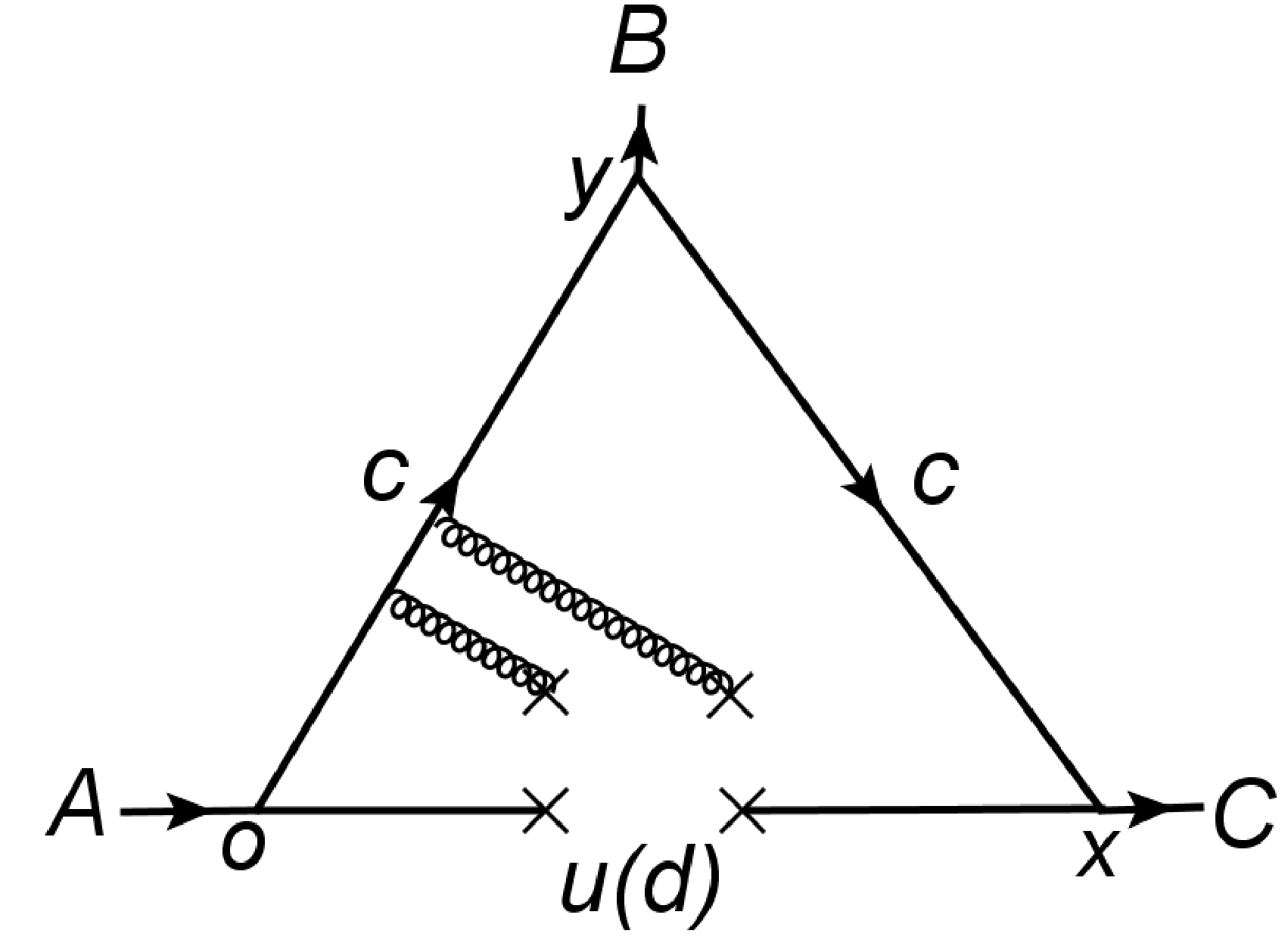}}
\subfigure[]{\includegraphics[height=3.2cm,width=3.5cm]{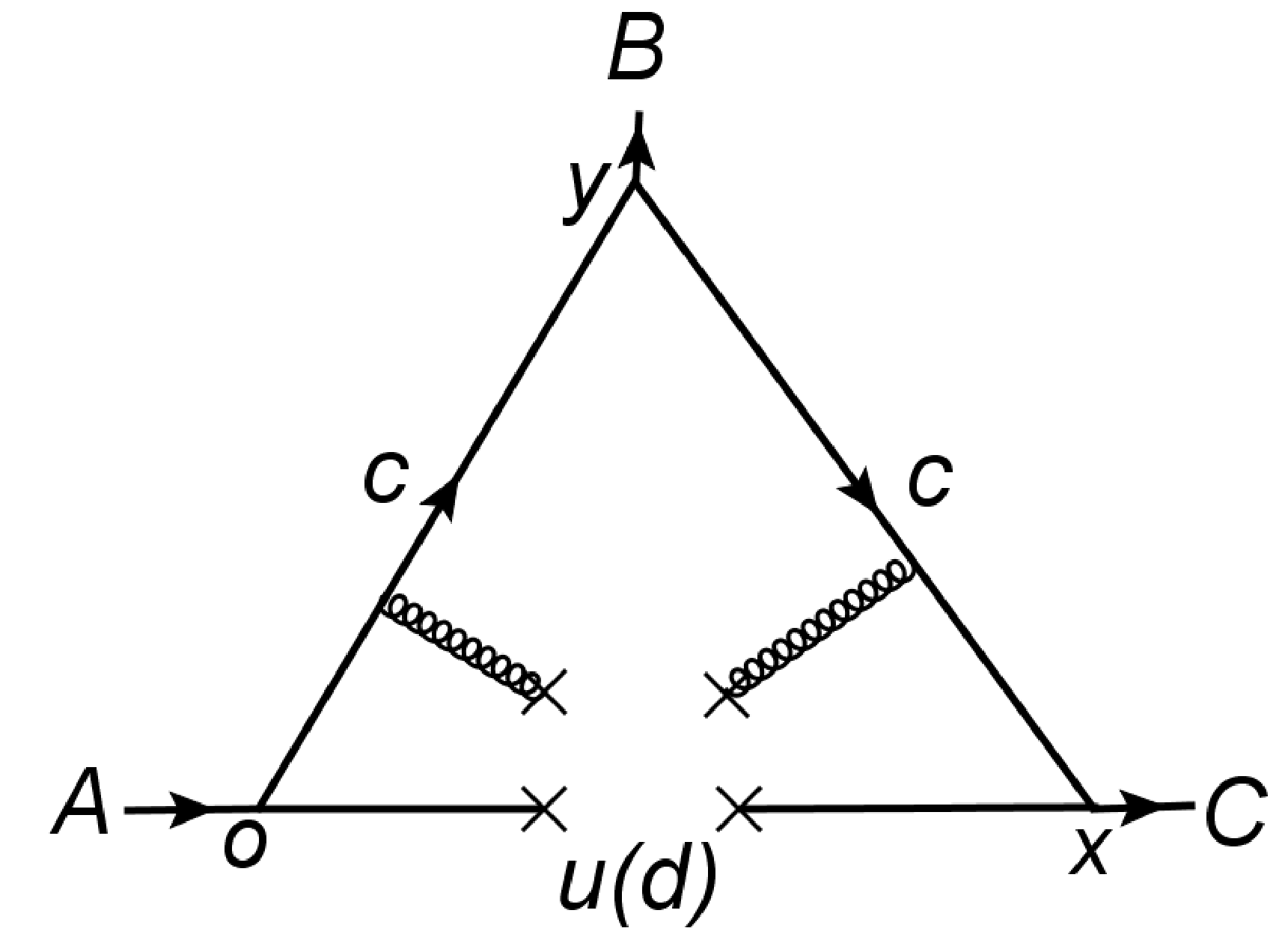}}
\caption{Contributions of the non-perturbative parts for $J/\psi(\eta_{c})$ off-shell.}
\label{fig:FM1}
\end{figure*}

\begin{figure*}[htbp]
\centering
\includegraphics[width=18cm]{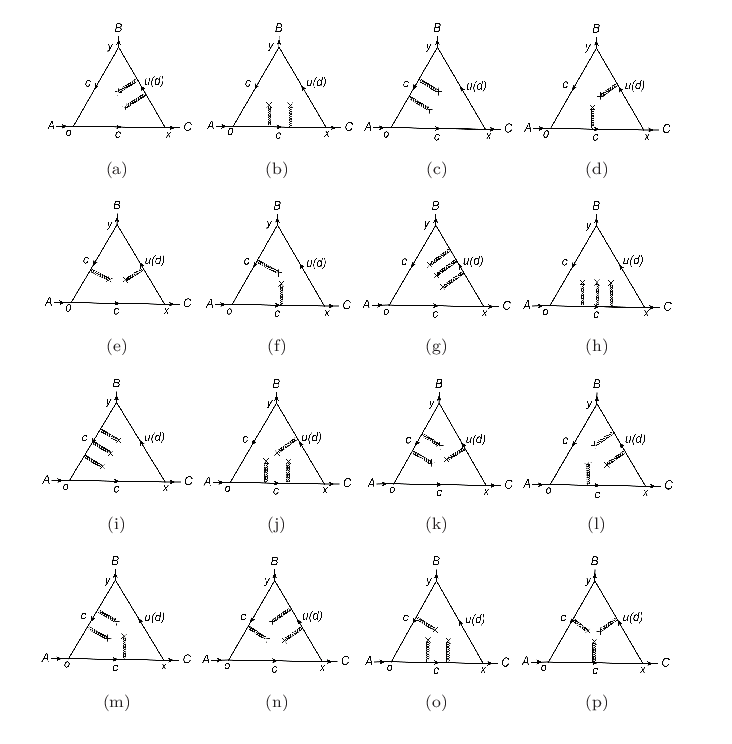}
\caption{Contributions of the non-perturbative parts for $D(D^{*})$ off-shell.}
\label{fig:FM2}
\end{figure*}

Besides of these above contributions, we also take into account the contributions from $\langle\overline{q}q\rangle$, $\langle\overline{q}g_{s}\sigma Gq\rangle$, and $\langle\overline{q}q\rangle\langle g_{s}^{2}G^{2}\rangle$. The feynman diagrams for these condensate terms can be classified into two groups which are shown in Figs. \ref{fig:FM1} and \ref{fig:FM2}.
As for the spectral density and contributions from $\langle\overline{q}q\rangle$, $\langle\overline{q}g_{s}\sigma Gq\rangle$, and $\langle\overline{q}q\rangle\langle g_{s}^{2}G^{2}\rangle$, only expressions for vertex $DDJ/\psi$ are shown in Appendixes A and B for simplicity.

We take the change of variables $p^{2}\to-P^{2}$, $p'^{2}\to-P'^{2}$ and $q^{2}\to-Q^{2}$ and perform double Borel transform\cite{Ioffe:1982ia,Ioffe:1982qb} to both the phenomenological and QCD sides. The variables $P^{2}$ and $P'^{2}$ will be replaced by $T_{1}^{2}$ and $T_{2}^{2}$, where $T_{1}$ and $T_{2}$ are the Borel parameters. In this article, we take $T^{2}=T_{1}^{2}$ and $T_{2}^{2}=kT_{1}^{2}=kT^{2}$, where $k$ is a constant related to meson mass. It takes different values for different vertices, and these values are represented in Table \ref{kE}.
Then, we match the phenomenological and the QCD sides using the quark-hardon duality, and the sum rules for the coupling constants are obtained. Finally, the momentum dependent coupling constants can be expressed as,
\begin{align}\label{eq:20}
g({Q^2}) = \frac{{ - \int\limits_{{s_1}}^{{s_0}} {\int\limits_{{u_1}}^{{u_0}} {\rho (s,u,{Q^2}){e^{ - s/{T^2}}}{e^{ - u/k{T^2}}}} dsdu + \mathscr B\mathscr B[{{\Pi }^{\mathrm{non - pert}}}]} }}{{\frac{E}{{(m_B^2 + {Q^2})}}{e^{ - m_A^2/{T^2}}}{e^{ - m_C^2/k{T^2}}}}}
\end{align}
Here, $\mathscr{BB}[~]$ stands for the double Borel transform, the factor $E$ has different expressions(see Table \ref{kE}) for different vertices. In Table \ref{kE}, $\Lambda(m_{J/\psi}^{2},m_{D^{*}}^{2},Q^{2})$ has the following form,
\begin{eqnarray}
\notag
\Lambda (m_{J/\psi }^2,m_{{D^*}}^2,{Q^2}) &&= \frac{{m_{{D^*}}^6 + m_{{D^*}}^4(m_{J/\psi }^2 + {Q^2})}}{{2m_{{D^*}}^2m_{J/\psi }^2{Q^2}}}- \frac{{{Q^4}}}{{2m_{{D^*}}^2m_{J/\psi }^2}}\\
\notag
&& - \frac{{m_{{D^*}}^2(m_{J/\psi }^4 - 10m_{J/\psi }^2{Q^2} + {Q^4})}}{{2m_{{D^*}}^2m_{J/\psi }^2{Q^2}}}\\
&& - \frac{{m_{J/\psi }^6 - 9m_{J/\psi }^4{Q^2} - 9m_{J/\psi }^2{Q^4}}}{{2m_{{D^*}}^2m_{J/\psi }^2{Q^2}}}
\end{eqnarray}
In order to eliminate the contributions of higher resonances and continuum states, the threshold parameters $s_{0}$ and $u_{0}$ in dispersion integral will be introduced in Eq. (\ref{eq:20}). They fulfill the relations, $m_{\mathrm{i}}^{2}<s_{0}<m'^{2}_{\mathrm{i}}$ and $m_{\mathrm{o}}^{2}<u_{0}<m'^{2}_{\mathrm{o}}$, where subscripts $i$ and $o$ represent incoming and outcoming mesons respectively. $m$ and $m'$ are the masses of the ground and first excited state of the mesons. They commonly have a relation $m'=m+\Delta$, where $\Delta$ is taken as a value of $0.4\sim0.6$ GeV\cite{Bracco:2011pg}.
\begin{table}[htbp]
\caption{The parameters $k$ and $E$ for different vertices and off-shell cases.}
\label{kE}
\begin{tabular}{c c c c}
\hline\hline
Vertices&off-shell&$k$&$E$ \\ \hline
\specialrule{0em}{1pt}{1pt}
\multirow{2}*{$DDJ/\psi$}&$J/\psi$ &$1$&$\frac{f_{D}^{2}m_{D}^{4}f_{J/\psi}m_{J/\psi}}{m_{c}^{2}}$  \\
\specialrule{0em}{1pt}{1pt}
~&$D$ &$\frac{m_{J/\psi}^{2}}{m_{D^{*}}^{2}}$&$\frac{2f_{D}^{2}m_{D}^{4}f_{J/\psi}m_{J/\psi}}{m_{c}^{2}}$\\
\specialrule{0em}{1pt}{1pt} \hline
\specialrule{0em}{1pt}{1pt}
\multirow{3}*{$DD^{*}J/\psi$}&$J/\psi$ &$\frac{m_{D^{*}}^{2}}{m_{D}^{2}}$&\multirow{3}*{$-\frac{f_{D}m_{D}^{2}f_{D^{*}}m_{D^{*}}f_{J/\psi}m_{J/\psi}}{m_{c}}$}  \\
\specialrule{0em}{1pt}{1pt}
~&$D$ &$\frac{m_{J/\psi}^{2}}{m_{D^{*}}^{2}}$&~   \\
\specialrule{0em}{1pt}{1pt}
~&$D^{*}$ &$\frac{m_{J/\psi}^{2}}{m_{D}^{2}}$&~   \\
\specialrule{0em}{1pt}{1pt} \hline
\specialrule{0em}{1pt}{1pt}
\multirow{2}*{$D^{*}D^{*}J/\psi$}&$J/\psi$ &$1$&$f_{D^{*}}^{2}m_{D^{*}}^{2}f_{J/\psi}m_{J/\psi}(5-\frac{Q^{2}}{2m_{D^{*}}^{2}})$  \\
\specialrule{0em}{1pt}{1pt}
~&$D^{*}$ &$\frac{m_{J/\psi}^{2}}{m_{D}^{2}}$&$f_{D^{*}}^{2}m_{D^{*}}^{2}f_{J/\psi}m_{J/\psi}\Lambda(m_{J/\psi}^{2},m_{D^{*}}^{2},Q^{2})$  \\
\specialrule{0em}{1pt}{1pt} \hline
\specialrule{0em}{1pt}{1pt}
\multirow{3}*{$DD^{*}\eta_{c}$}&$\eta_{c}$ &$\frac{m_{D^{*}}^{2}}{m_{D}^{2}}$&\multirow{2}*{$-\frac{f_{D}m_{D}^{2}f_{D^{*}}m_{D^{*}}f_{\eta_{c}}m_{\eta_{c}}^{2}}{m_{c}^{2}}$}  \\
\specialrule{0em}{1pt}{1pt}
~&$D$ &$\frac{m_{\eta_{c}}^{2}}{m_{D^{*}}^{2}}$&~   \\
\specialrule{0em}{1pt}{1pt}
~&$D^{*}$ &$\frac{m_{\eta_{c}}^{2}}{m_{D}^{2}}$&$\frac{f_{D}m_{D}^{2}f_{D^{*}}m_{D^{*}}f_{\eta_{c}}m_{\eta_{c}}^{2}(m_{\eta_{c}}^{2}+m_{D^{*}}^{2}-m_{D}^{2})}{2m_{c}^{2}m_{D^{*}}^{2}}$   \\
\specialrule{0em}{1pt}{1pt} \hline
\specialrule{0em}{1pt}{1pt}
\multirow{2}*{$D^{*}D^{*}\eta_{c}$}&$\eta_{c}$ &$1$&\multirow{2}*{$-\frac{f_{D^{*}}^{2}m_{D^{*}}^{2}f_{\eta_{c}}m_{\eta_{c}}^{2}}{2m_{c}^{2}}$}  \\
\specialrule{0em}{1pt}{1pt}
~&$D^{*}$ &$\frac{m_{\eta_{c}}^{2}}{m_{D^{*}}^{2}}$&~   \\
\specialrule{0em}{1pt}{1pt} \hline\hline
\end{tabular}
\end{table}
To obtain the final results of strong coupling constant, it is necessary to extrapolate $g({Q^2})$ into time-like regions $(Q^{2}<0)$. This process is realized by fitting the $g(Q^{2})$ with appropriate analytical functions and by setting the intermediate meson on-shell($Q^2=-m_{\mathrm{on-shell}}^{2}$).

\section{Numerical results and Discussions}\label{sec3}

The hadronic parameters used in present work are
taken as $m_{D}=1.86$ GeV\cite{ParticleDataGroup:2022pth}, $m_{D^{*}}=2.01$ GeV\cite{ParticleDataGroup:2022pth}, $m_{J/\psi}=3.09$ GeV\cite{ParticleDataGroup:2022pth}, $m_{\eta_{c}}=2.98$ GeV\cite{ParticleDataGroup:2022pth}, $m_{u(d)}=0.006\pm0.001$ GeV\cite{ParticleDataGroup:2022pth}, $m_{c}=1.275\pm0.025$ GeV\cite{ParticleDataGroup:2022pth}, $f_{D}=0.210\pm0.011$ GeV\cite{Wang:2015mxa}, $f_{D_{*}}=0.263\pm0.021$ GeV\cite{Wang:2015mxa}, $f_{J/\psi}=0.418$ GeV\cite{Becirevic:2013bsa}, $f_{\eta_{c}}=0.387$ GeV\cite{Becirevic:2013bsa}. The vacuum condensates are taken as $\langle\overline{q}q\rangle=-(0.23\pm0.01)^{3}$ GeV$^{3}$\cite{ParticleDataGroup:2022pth}, $\langle\overline{q}g_{s}\sigma Gq\rangle=m_{0}^{2}\langle\overline{q}q\rangle$\cite{ParticleDataGroup:2022pth}, $m_{0}^{2}=0.8\pm0.1$ GeV$^2$\cite{Narison:2010cg,Narison:2011xe,Narison:2011rn}, $\langle g_{s}^{2}G^{2}\rangle=0.88\pm0.15$ GeV$^{4}$\cite{Narison:2010cg,Narison:2011xe,Narison:2011rn}, $\langle f^{3}G^{3}\rangle=(8.8\pm5.5)$ GeV$^{2}\langle g_{s}^{2}G^{2}\rangle$\cite{Narison:2010cg,Narison:2011xe,Narison:2011rn}. The threshold parameters in Eq. (\ref{eq:20}) are defined as $s_{0}=(m_{\mathrm{i}}+\Delta_{\mathrm{i}})^{2}$ and $u_{0}=(m_{\mathrm{o}}+\Delta_{\mathrm{o}})^{2}$. We uniformly take $\Delta_{\mathrm{i}}=\Delta_{\mathrm{o}}=$0.4, 0.5 and 0.6 GeV, where 0.5 GeV is used to obtain the central values of coupling constants, and 0.4, 0.6 GeV are for lower and upper bounds of the results, respectively.

To ensure the reliability of the final results, two conditions should be satisfied, which are the pole dominance and convergence of operator product expansion(OPE). We firstly write,
\begin{eqnarray}
\notag
\Pi^{\mathrm{OPE}}_{\mathrm{pole}}(T^{2})=-\int_{s_{1}}^{s_{0}}\int_{u_{1}}^{u_{0}}\rho^{\mathrm{OPE}}(s,u,Q^2)e^{-\frac{s}{T^{2}}}e^{-\frac{u}{kT^{2}}}dsdu \\
\Pi^{\mathrm{OPE}}_{\mathrm{cont}}(T^{2})=-\int_{s_{0}}^{\infty}\int_{u_{0}}^{\infty}\rho^{\mathrm{OPE}}(s,u,Q^2)e^{-\frac{s}{T^{2}}}e^{-\frac{u}{kT^{2}}}dsdu
\end{eqnarray}
Then, the pole and continuum contribution can be defined as\cite{Bracco:2011pg},
\begin{eqnarray}
\notag
\mathrm{Pole}=\frac{\Pi^{\mathrm{OPE}}_{\mathrm{pole}}(T^{2})}{\Pi^{\mathrm{OPE}}_{\mathrm{pole}}(T^{2})+\Pi^{\mathrm{OPE}}_{\mathrm{cont}}(T^{2})} \\
\mathrm{Continuum}=\frac{\Pi^{\mathrm{OPE}}_{\mathrm{cont}}(T^{2})}{\Pi^{\mathrm{OPE}}_{\mathrm{pole}}(T^{2})+\Pi^{\mathrm{OPE}}_{\mathrm{cont}}(T^{2})}
\end{eqnarray}

Fixing $Q^{2}=3$ GeV$^{2}$ in Eq. (\ref{eq:20}), we plot the contributions of pole and continuum in Fig. \ref{PC}.
Besides, the contributions of pertubative part and different vacuum condensate terms are shown in Fig. \ref{BW}.
From Fig. \ref{BW}, we find good stability of the results. This stable region is called the Borel platform. The appearance of Borel platform indicates the condition of convergence of OPE is satisfied.

Then, we choose central value of Borel parameter(denoted as $T_{0}^{2}$ in Table~\ref{BTI}) in the Borel platform to meet the condition of pole dominance($>40\%$) and get the coupling constants. Finally, by taking different values of $Q^{2}$, we obtain coupling constant $g(Q^{2})$. The Borel platform, $T_{0}^{2}$, pole contributions at $T_{0}^{2}$ and the range of $Q^{2}$ which is used to get coupling constants, are shown in Table~\ref{BTI}.
\begin{table}[htbp]
\begin{ruledtabular}\caption{The Borel platform, $T_{0}^{2}$, pole contributions(Pole) and the range of $Q^{2}$ for all vertices. Except for pole contributions, all values are in units of GeV$^{2}$.}
\label{BTI}
\begin{tabular}{c c c c c c }
Vertices&off-shell&Borel platform&$T_{0}^{2}$&Pole&$Q^{2}$ \\ \hline
\multirow{2}*{$DDJ/\psi$}&$J/\psi$ &$4.5-6.5$&$5.5$&$40.51\%$&$1-8$  \\
~&$D$ &$4-6$&$4.8$&$40.2\%$&$4-10$   \\  \hline
\multirow{3}*{$DD^{*}J/\psi$}&$J/\psi$ &$5-7$&$6$&$40.21\%$&$3.5-9.5$  \\
~&$D$ &$5.5-7.5$&$6.5$&$42.5\%$&$3-9$   \\
~&$D^{*}$ &$4.5-6.5$&$5.5$&$42.16\%$&$3-9$   \\ \hline
\multirow{2}*{$D^{*}D^{*}J/\psi$}&$J/\psi$ &$4.5-6.5$&$5.8$&$40.2\%$&$1-7$  \\
~&$D^{*}$ &$4-6$&$5$&$42.63\%$&$5-11$   \\  \hline
\multirow{3}*{$DD^{*}\eta_{c}$}&$\eta_{c}$ &$4-6$&$4.8$&$40.51\%$&$2-8$  \\
~&$D$ &$3.5-5.5$&$4.5$&$40.03\%$&$3-9$   \\
~&$D^{*}$ &$2-4$&$2.5$&$40.11\%$&$3-9$   \\ \hline
\multirow{2}*{$D^{*}D^{*}\eta_{c}$}&$\eta_{c}$ &$6-8$&$7$&$41.05\%$&$3-9$  \\
~&$D^{*}$ &$5-7$&$6$&$43.23\%$&$3-9$   \\
\end{tabular}
\end{ruledtabular}
\end{table}

\begin{figure}[H]
\centering
\includegraphics[width=5cm]{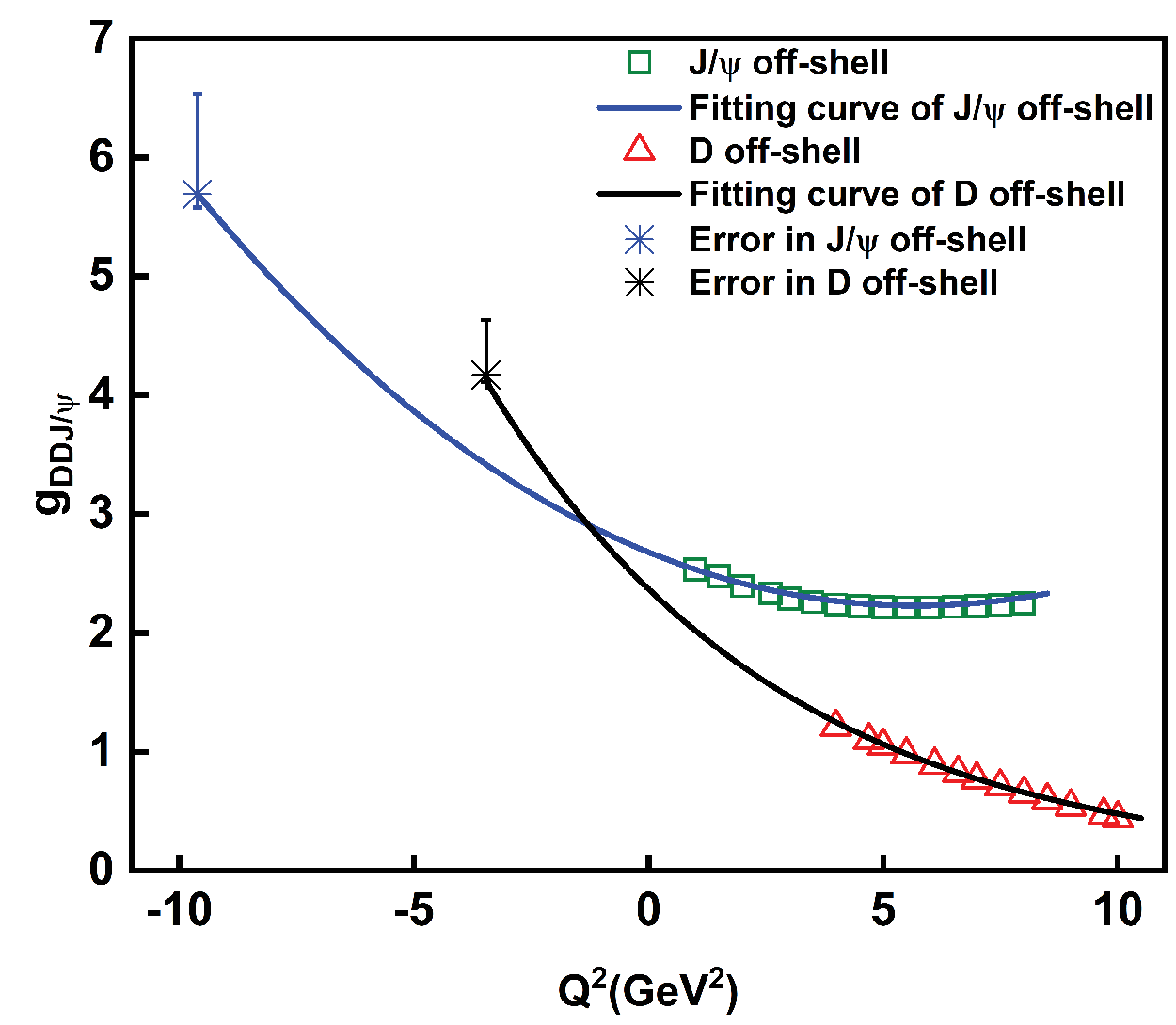}
\caption{The fitting curves of coupling constants for $DDJ/\psi$ vertex. In this figure, different off-shell cases($J/\psi$ and $D$ mesons) are considered.}
\label{fig:FF1}
\end{figure}

\begin{figure}[H]
\centering
\includegraphics[width=5cm]{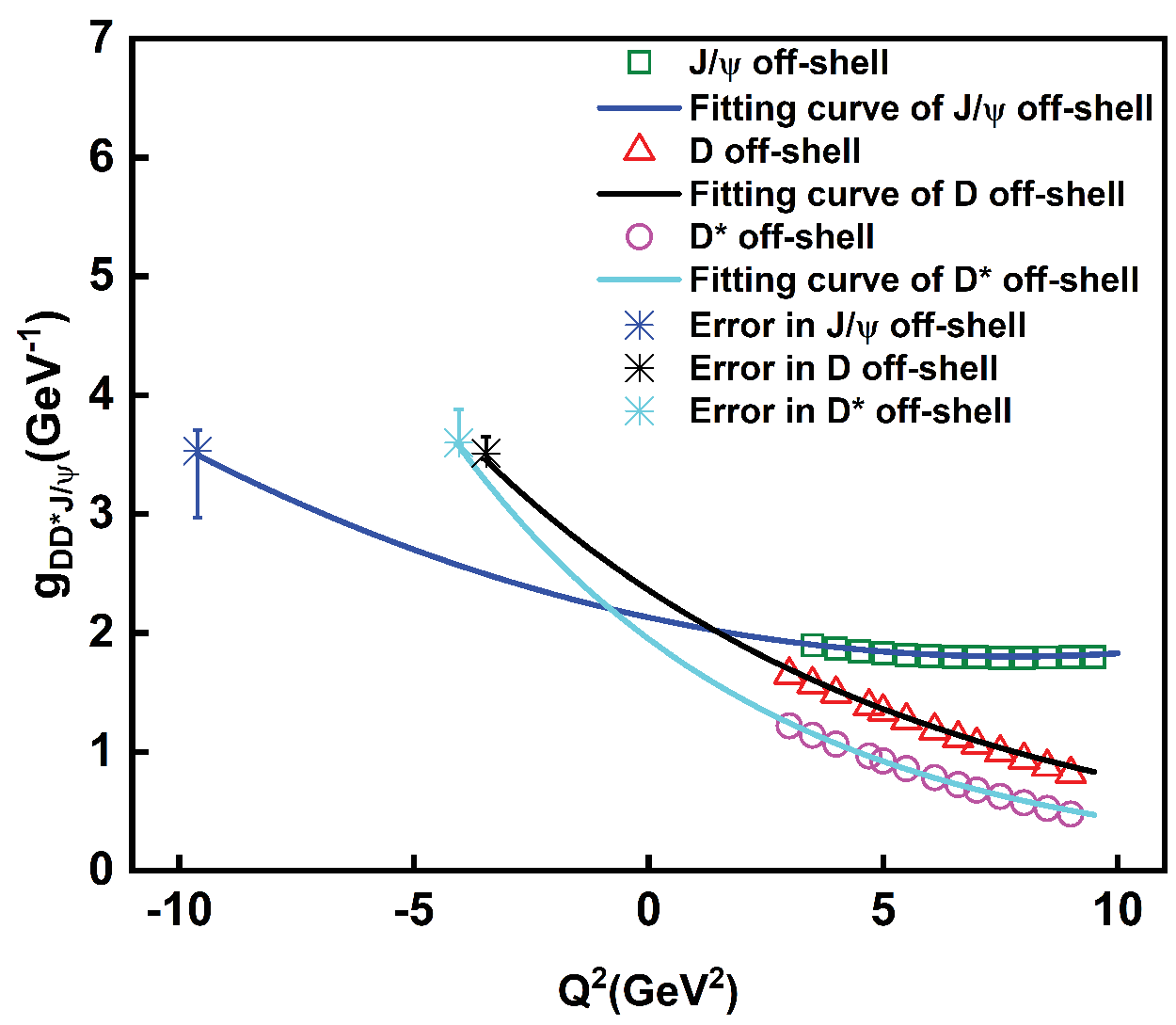}
\caption{The fitting curves of coupling constants for $DD^{*}J/\psi$ vertex. In this figure, different off-shell cases($J/\psi$, $D$ and $D*$ mesons) are considered.}
\label{fig:FF2}
\end{figure}

\begin{figure}[H]
\centering
\includegraphics[width=5cm]{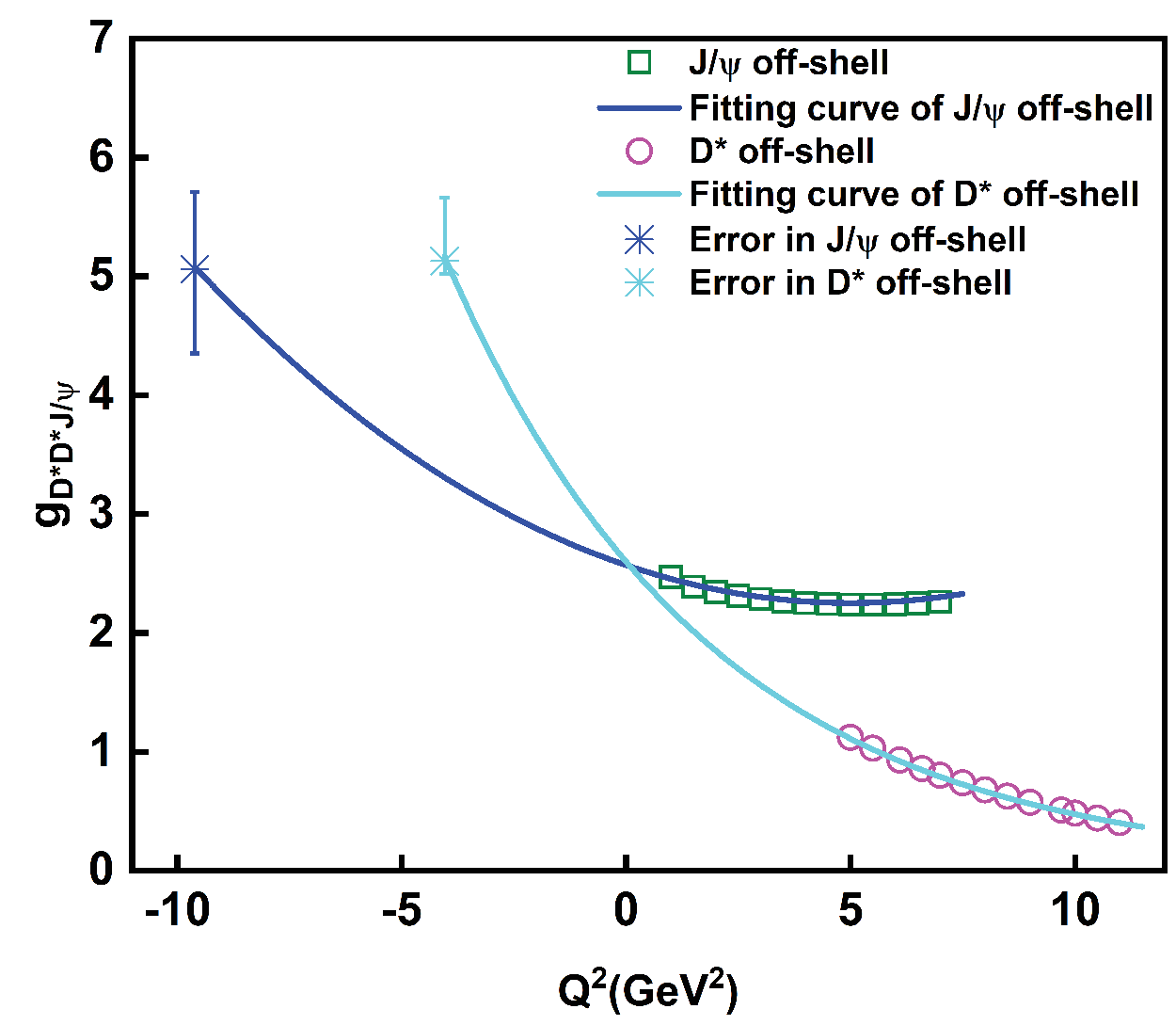}
\caption{The fitting curves of coupling constants for $D^{*}D^{*}J/\psi$ vertex. In this figure, different off-shell cases($J/\psi$ and $D*$ mesons) are considered.}
\label{fig:FF3}
\end{figure}

\begin{figure}[H]
\centering
\includegraphics[width=5cm]{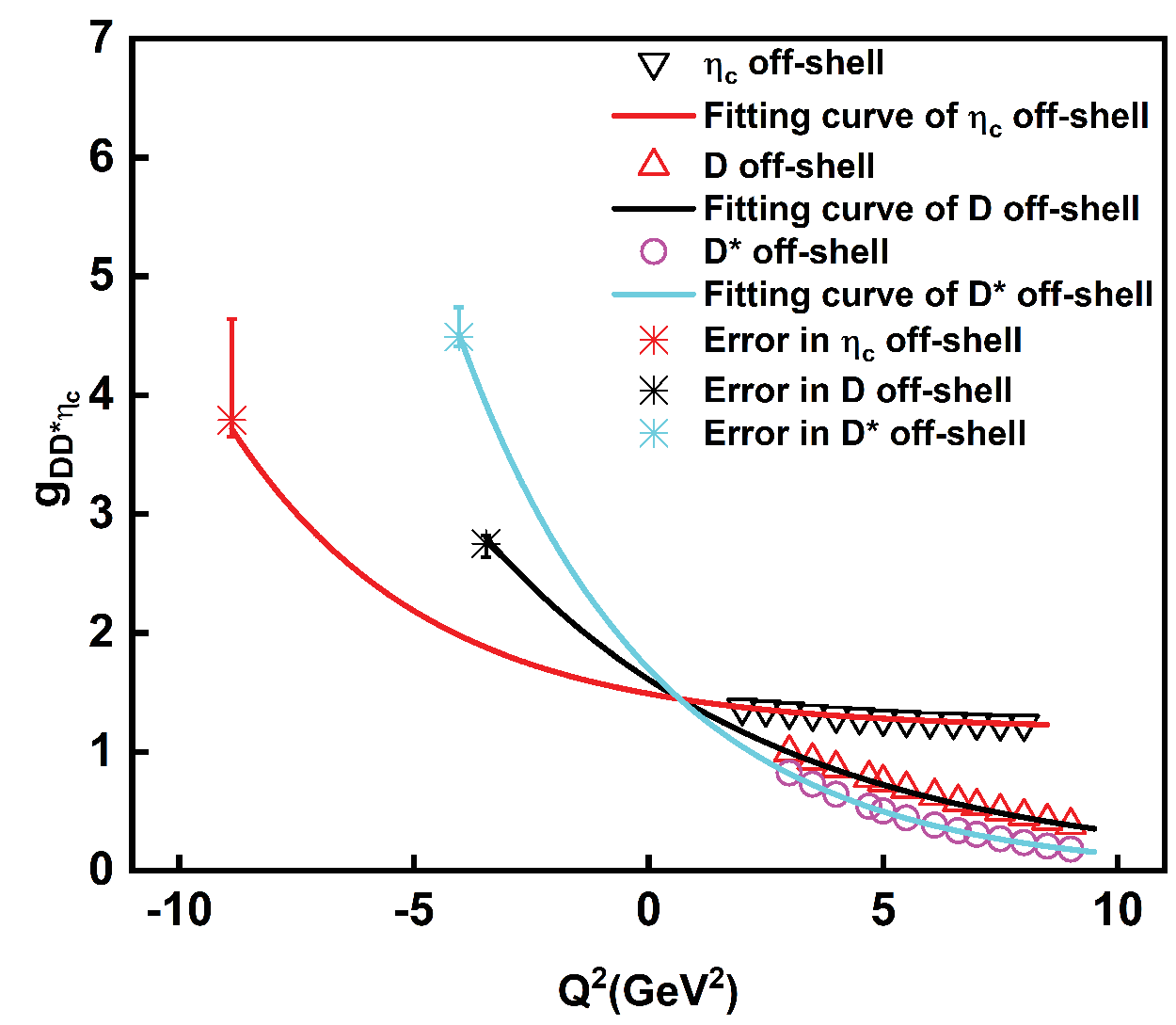}
\caption{The fitting curves of coupling constants for $DD^{*}\eta_{c}$ vertex. In this figure, different off-shell cases($\eta_{c}$, $D$ and $D*$ mesons) are considered.}
\label{fig:FF4}
\end{figure}

\begin{figure}[H]
\centering
\includegraphics[width=5cm]{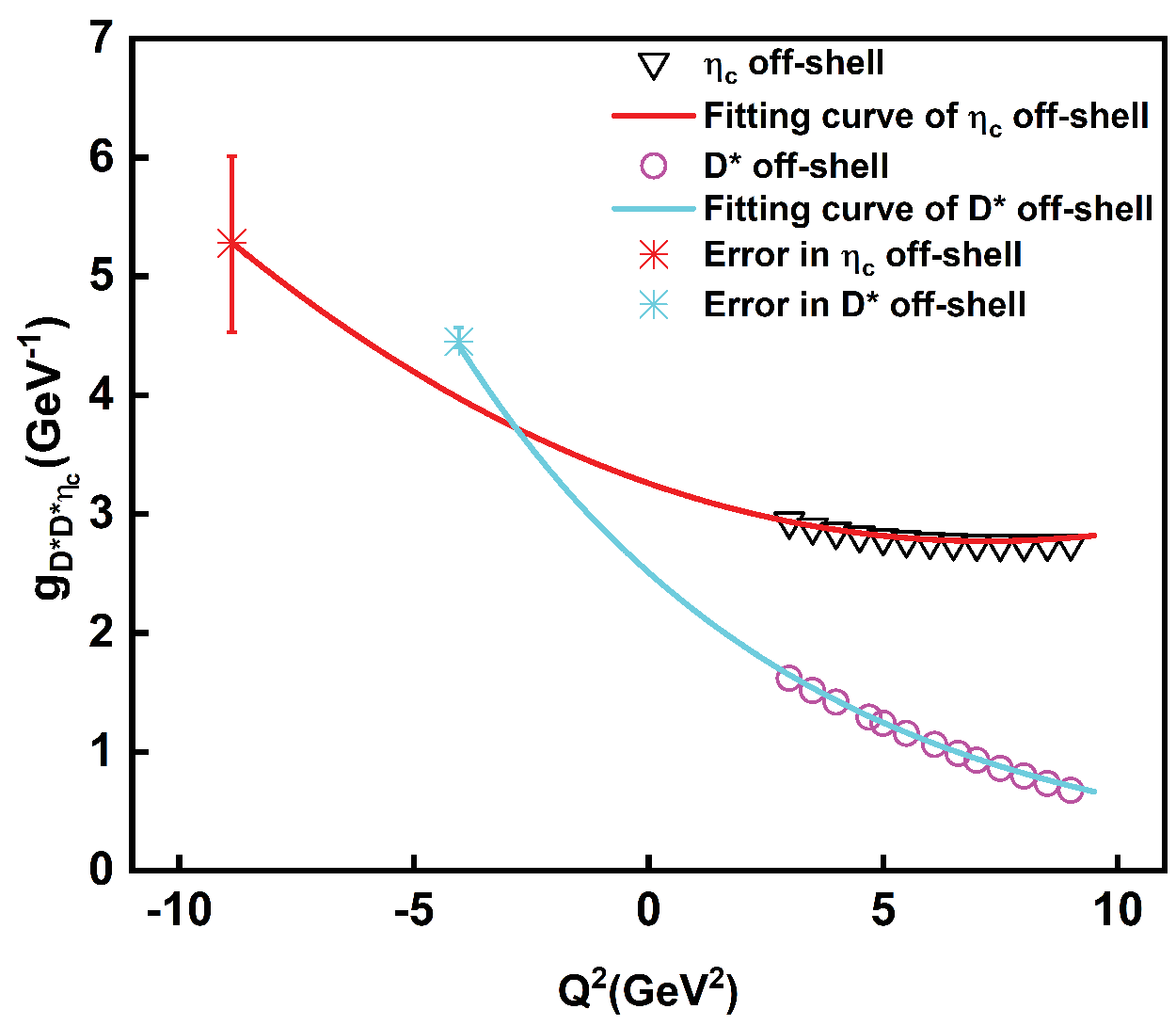}
\caption{The fitting curves of coupling constants for $D^{*}D^{*}\eta_{c}$ vertex. In this figure, different off-shell cases($\eta_{c}$ and $D*$ mesons) are considered.}
\label{fig:FF5}
\end{figure}

The momentum dependent strong coupling constants can be fitted uniformly by following analytical function,
\begin{eqnarray}\label{eq:26}
g(Q^{2})=A_{1}e^{-B_{1}Q^{2}}+C_{1}+C_{2}Q^{4}
\end{eqnarray}
where the parameters $A_{1},B_{1},C_{1}$ and $C_{2}$ are shown in Table~\ref{FFSC}. The fitting diagrams of strong coupling constants for each vertex are shown in Figs. \ref{fig:FF1}-\ref{fig:FF5}.
Then the $g(Q^{2})$ is extrapolated into the time-like region $(Q^{2}<0)$ by Eq. (\ref{eq:26}), and on-shell condition is satisfied by setting $Q^{2}=-m_{\mathrm{on-shell}}^{2}$. For each vertex, we finally obtain the values of strong coupling constants for different off-shell cases. The results are shown in the last column of Table~\ref{FFSC}.
\begin{table}[htbp]
\begin{ruledtabular}\caption{The parameters for the analysis function and the on-shell values of strong coupling constants for different vertices and off-shell cases.}
\label{FFSC}
\begin{tabular}{l l l l l l l }
Vertex&off-shell&$A_{1}$&$B_{1}$&$C_{1}$&$C_{2}$&$g(Q^{2}=-m_{\mathrm{on-shell}}^{2})$ \\ \hline
\multirow{2}*{$DDJ/\psi$}&$J/\psi$ &$2.68$&$0.06$&$0$&$0.010$&$5.83^{+0.70}_{-0.25}$  \\
~&$D$ &$2.37$&$0.16$&$0$&$0$&$4.17^{+0.46}_{-0.06}$   \\  \hline
\multirow{3}*{$DD^{*}J/\psi$}&$J/\psi$ &$2.13$&$0.04$&$0$&$0.004$&$3.53^{+0.18}_{-0.56}$ GeV$^{-1}$  \\
~&$D$ &$2.36$&$0.11$&$0$&$0$&$3.51^{+0.14}_{-0.05}$ GeV$^{-1}$   \\
~&$D^{*}$ &$1.95$&$0.15$&$0$&$0$&$3.60^{+0.28}_{-0.01}$ GeV$^{-1}$   \\ \hline
\multirow{2}*{$D^{*}D^{*}J/\psi$}&$J/\psi$ &$2.57$&$0.05$&$0$&$0.01$&$5.06^{+0.65}_{-0.71}$  \\
~&$D^{*}$ &$2.60$&$0.17$&$0$&$0$&$5.13^{+0.53}_{-0.11}$   \\  \hline
\multirow{3}*{$DD^{*}\eta_{c}$}&$\eta_{c}$ &$0.30$&$0.24$&$1.19$&$0$&$3.79^{+0.85}_{-0.14}$  \\
~&$D$ &$1.61$&$0.16$&$0$&$0$&$2.75^{+0.07}_{-0.11}$   \\
~&$D^{*}$ &$1.72$&$0.24$&$-0.02$&$0$&$4.49^{+0.25}_{-0.08}$   \\ \hline
\multirow{2}*{$D^{*}D^{*}\eta_{c}$}&$\eta_{c}$ &$3.26$&$0.042$&$0$&$0.007$&$5.28^{+0.73}_{-0.75}$ GeV$^{-1}$  \\
~&$D^{*}$ &$2.51$&$0.14$&$0$&$0$&$4.45^{+0.12}_{-0.05}$ GeV$^{-1}$   \\
\end{tabular}
\end{ruledtabular}
\end{table}
For each vertex, the strong coupling constants which are obtained by considering different off-shell cases, should be equal to each other. For $g_{DD^{*}J/\psi}$ as an example, it is shown in Table~\ref{FFSC} that the results of $g_{DD^{*}J/\psi}^{D}$, $g_{DD^{*}J/\psi}^{D^{*}}$ and $g_{DD^{*}J/\psi}^{J/\psi}$ for different off-shell cases are consistent with each other.
Thus, by taking average value of these results, we can obtain the strong coupling constants for each vertex. All of the final results are collected in Table~\ref{FFFF}, where we can see that there is a significant difference between the results of QCDSR and those of other methods. In Refs.\cite{Bracco:2011pg,Rodrigues:2017qsm}, the authors also carried out the same work with the same method where the OPE was truncated to 3 dimensions. In the present work, the OPE was truncated at dimension of 7. Considering the uncertainties, we can see that our results are compatible with those in Refs.\cite{Bracco:2011pg,Rodrigues:2017qsm}, although there are slightly differences between them. This closeness of the results is a good manifestation that effectively demonstrates the convergence of the OPE.
\begin{table*}[htbp]
\begin{ruledtabular}\caption{The strong coupling constants for all vertices.}
\label{FFFF}
\begin{tabular}{c c c c c }
Vertices&Present work&QCDSR &VMD&Other models \\  \hline
$DDJ/\psi$&$5.01^{+0.58}_{-0.16}$&$5.8\pm0.9$\cite{Bracco:2011pg}&7.64\cite{Lin:1999ad}&$8.0\pm0.5$\cite{Deandrea:2003pv}\\
$DD^{*}J/\psi$&$3.55^{+0.20}_{-0.21}$ GeV$^{-1}$&$4.0\pm0.6$ GeV$^{-1}$\cite{Bracco:2011pg}&$8.0\pm0.6$ GeV$^{-1}$\cite{Oh:2000qr}&$4.05\pm0.25$ GeV$^{-1}$\cite{Deandrea:2003pv} \\
$D^{*}D^{*}J/\psi$&$5.10^{+0.59}_{-0.43}$&$6.2\pm0.9$\cite{Bracco:2011pg}&7.64\cite{Lin:1999ad}&$8.0\pm0.5$\cite{Deandrea:2003pv} \\
$DD^{*}\eta_{c}$&$3.68^{+0.39}_{-0.11}$&$5.23^{+1.80}_{-1.38}$\cite{Rodrigues:2017qsm}&7.68\cite{Wang:2012wj}&$15.51\pm0.45$\cite{Lucha:2015dda} \\
$D^{*}D^{*}\eta_{c}$&$4.87^{+0.42}_{-0.40}$ GeV$^{-1}$&-&-&$9.76\pm0.32$ GeV$^{-1}$\cite{Lucha:2015dda} \\
\end{tabular}
\end{ruledtabular}
\end{table*}

According to SU(4) symmetry, the strong coupling constants of vertices $DDJ/\psi$ and $D^{*}D^{*}J/\psi$ statisfy the relation, $\frac{g_{DDJ/\psi}}{g_{D^{*}D^{*}J/\psi}}=1$\cite{Bracco:2011pg}. And there are following relations, $\frac{g_{DDJ/\psi}}{g_{DD^{*}J/\psi}}=m_{D}$\cite{Bracco:2011pg}, $\frac{g_{DD^{*}\eta_{c}}}{g_{D^{*}D^{*}\eta_{c}}}=\frac{m_{D}}{2}$, $g_{DD^{*}\eta_{c}}=g_{2}\sqrt{m_{\eta_{c}}}m_{D}$, and $g_{2}=2.36$GeV$^{-3/2}$ in the heavy quark effective theory\cite{Lin:2017mtz}. However, in charmed sector, the SU(4) and heavy-quark spin symmetry is only approximate, these relations are violated\cite{Bracco:2011pg}. In our calculations, this violation is proved by the following relations of strong coupling constants, which are  $\frac{g_{DDJ/\psi}}{g_{D^{*}D^{*}J/\psi}}\approx0.98<1$, $\frac{g_{DDJ/\psi}}{g_{DD^{*}J/\psi}}\approx1.41<m_{D}$, $\frac{g_{DD^{*}\eta_{c}}}{g_{D^{*}D^{*}\eta_{c}}}\approx0.76<\frac{m_{D}}{2}$ and $g_{DD^{*}\eta_{c}}=3.68<g_{2}\sqrt{m_{\eta_{c}}}m_{D}$.

\section{Conclusions}\label{sec4}

In this work, we analyze the strong vertices $DDJ/\psi$, $DD^{*}J/\psi$, $D^{*}D^{*}J/\psi$, $DD^{*}\eta_{c}$, $D^{*}D^{*}\eta_{c}$ by different QCD sum rules, where all off-shell cases are considered for each vertex. Under this physical sketch, the momentum dependent strong coupling constants are fristly calculated in the space-like ($Q^{2}>0$) regions, and then are fitted into appropriate functions. By extrapolating these functions into the time-like ($Q^{2}<0$) regions and taking $Q^{2}=-m^{2}_{\mathrm{on-shell}}$, we obtain the strong coupling constants. For each vertex, we take the average value of the strong coupling constants for different off-shell cases as the final results. These coupling constants are important in describing the dynamical behaviours of hadrons, which can be used to analyze the production processes of the exotic hadrons.

\section*{Acknowledgements}

This project is supported by National Natural Science Foundation, Grant Number 12175068 and Natural Science Foundation of HeBei Province, Grant Number A2018502124.

\begin{widetext}

\begin{large}
\textbf{Appendix A: Full expressions of the perturbative, $\langle g_{s}^{2}G^{2}\rangle$ and $\langle f^{3}G^{3}\rangle$ spectral density for $J/\psi$ and $D$ off-shell cases.}\label{B}
\end{large}

\begin{eqnarray}
\notag
{\rho ^{pert(J/\psi )}}&& =  - \frac{3}{{4{\pi ^2}{{[{{({Q^2} + s + u)}^2} - 4su]}^{5/2}}}}\{  - m_c^6{Q^2}({Q^2} + s - u) + m_c^4{Q^2}[ - 2{Q^4}- {Q^2}(4s + u)\\
\notag
&& - 2{s^2} + su + {u^2}] + m_c^2[{Q^6}(s + u) + {Q^4}(3{s^2} + 2{u^2}) + {Q^2}(3{s^3} - 4{s^2}u + {u^3})\\
&& + s{(s - u)^3}] + su[{Q^4}s + {Q^2}(2{s^2} - su - {u^2}) + {(s - u)^3}]\}
\end{eqnarray}
\begin{eqnarray}
\notag
{\rho ^{\left\langle {g_s^2{G^2}} \right\rangle (J/\psi )}}&& = \frac{{\left\langle {g_s^2{G^2}} \right\rangle }}{{96{\pi ^2}{{[{{({Q^2} + s + u)}^2} - 4su]}^{7/2}}}}\{ 12m_c^4[2{Q^4}s - {Q^2}({s^2} - 4su + 5{u^2}) - 3{s^3}\\
\notag
&& + {s^2}u - 3s{u^2} + 5{u^3}] + 6m_c^2[2{Q^8} + 2{Q^6}(4s + u) + {Q^4}(7{s^2} + 2su - 7{u^2})\\
\notag
&& - 2{Q^2}({s^3} + {s^2}u - s{u^2} + {u^3}) - 3{s^4} - 2{s^3}u - 2{s^2}{u^2} + 2s{u^3} + 5{u^4}] + 5{Q^{10}} \\
\notag
&&+ {Q^8}(13s + 15u) + 2{Q^6}({s^2} + 3su + 14{u^2}) - 2{Q^4}(11{s^3} + 6{s^2}u - 5s{u^2} - 22{u^3})\\
&& + {Q^2}( - 23{s^4} + 18{s^3}u - 18s{u^3} + 39{u^4}) - {(s - u)^2}(7{s^3} - 7{s^2}u + 9s{u^2} - 13{u^3})\}
\end{eqnarray}
\begin{eqnarray}
\notag
{\rho ^{\left\langle {{f^3}{G^2}} \right\rangle (J/\psi )}} &&=  - \frac{{\left\langle {{f^3}{G^2}} \right\rangle }}{{192{\pi ^2}{{[{{({Q^2} + s + u)}^2} - 4su]}^{9/2}}}}\{ 126m_c^5{Q^6}({Q^2} + s - u) - 9m_c^4[35{Q^8}\\
\notag
&& + {Q^6}(36s - 35u) - 3{Q^4}({s^2} - su + 6{u^2}) - {Q^2}(2{s^3} + 3{s^2}u + 33s{u^2} - 48{u^3})\\
\notag
&& + {(s - u)^2}(2{s^2} + su - 18{u^2})] + 18m_c^3{Q^4}[{Q^6} - 2{Q^4}(2s + 3u) - {Q^2}(11{s^2} - 12su + {u^2})\\
\notag
&& - 6{(s - u)^3}] + m_c^2[ - 142{Q^{10}} - 16{Q^8}(4s - 7u) + 3{Q^6}(126{s^2} - 134su + 89{u^2})\\
\notag
&& + {Q^4}(386{s^3} - 558{s^2}u + 1359s{u^2} - 527{u^3}) + {Q^2}(92{s^4} + 100{s^3}u + 873{s^2}{u^2}\\
\notag
&& - 874s{u^3} - 571{u^4}) + {(s - u)^2}(6{s^3} + 12{s^2}u + 233s{u^2} - 31{u^3})] - 9{m_c}{Q^2}[{Q^8}(s + u)\\
\notag
&& + {Q^6}({s^2} - 6su - {u^2}) - 3{Q^4}({s^3} + {s^2}u - 3s{u^2} + {u^3}) - {Q^2}{(s - u)^3}(5s + u) - 2{(s - u)^5}]\\
\notag
&& + {Q^{10}}(77s + 52u) + {Q^8}(229{s^2} - 35su + 124{u^2}) + {Q^6}(218{s^3} - 115{s^2}u + 405s{u^2} + 25{u^3})\\
\notag
&& + {Q^4}(50{s^4} + 125{s^3}u + 357{s^2}{u^2} + 58s{u^3} - 122{u^4}) - {Q^2}(23{s^5} - 159{s^4}u + 97{s^3}{u^2}\\
&& - 311{s^2}{u^3} + 317s{u^4} + 83{u^5}) - {(s - u)^2}(7{s^4} + 8{s^3}u - 24{s^2}{u^2} - 144s{u^3} + 8{u^4})\}
\end{eqnarray}
\begin{eqnarray}
\notag
{\rho ^{pert(D)}} &&=  - \frac{3}{{4{\pi ^2}{{[{{({Q^2} + s + u)}^2} - 4su]}^{5/2}}}}\{  - 2m_c^6{u^2} + m_c^4u[{Q^4} + {Q^2}(2s + 5u) + {s^2} - 5su + 4{u^2}]\\
\notag
&& + m_c^2[{Q^8} + 4{Q^6}(s + u) + {Q^4}(6{s^2} + 4su + 5{u^2}) + 2{Q^2}(2{s^3} - 2{s^2}u + {u^3}) + s(s - 2u){(s - u)^2}]\\
&& + {Q^2}su[{Q^4} + {Q^2}(2s + u) + s(s - u)]\}
\end{eqnarray}
\begin{eqnarray}
\notag
{\rho ^{\left\langle {g_s^2{G^2}} \right\rangle (D)}}&& = \frac{{\left\langle {g_s^2{G^2}} \right\rangle }}{{48{\pi ^2}{{[{{({Q^2} + s + u)}^2} - 4su]}^{7/2}}}}\{ 12m_c^4[{Q^6} + {Q^4}(s - 3u) + {Q^2}( - {s^2} - 4su + {u^2})\\
\notag
&& - s({s^2} + 3su + {u^2})] - 6m_c^2[{Q^8} - 2{Q^6}u - 2{Q^4}{s^2} + {Q^2}{u^2}(5u - 2s) + {s^4} + 2{s^3}u - 5s{u^3} \\
\notag
&& + 2{u^4}]+ 2{Q^{10}} + 3{Q^8}(2s + u) + 2{Q^6}(2{s^2} + {u^2}) + {Q^4}( - 4{s^3} - 2{s^2}u + 6s{u^2} + 8{u^3}) \\
&& - 2{Q^2}(3{s^4} - 4{s^3}u + 5{s^2}{u^2} + 2s{u^3} - 6{u^4}) - {(s - u)^3}(2{s^2} - su + 5{u^2})\}
\end{eqnarray}
\begin{eqnarray}
\notag
{\rho ^{\left\langle {{f^3}{G^2}} \right\rangle (D)}}&& = \frac{{\left\langle {{f^3}{G^2}} \right\rangle }}{{192{\pi ^2}{{[{{({Q^2} + s + u)}^2} - 4su]}^{9/2}}}}\{ 9m_c^4[7{Q^8} + {Q^6}(23s - 32u) + 3{Q^4}(9{s^2} - 9su\\
\notag
&& + 14{u^2}) + {Q^2}(13{s^3} + 27{s^2}u + 57s{u^2} - 2{u^3}) + 2{s^4} + 22{s^3}u + 42{s^2}{u^2} + 17s{u^3} - 83{u^4}]\\
\notag
&& + 2m_c^2[15{Q^{10}} + {Q^8}(91s + 109u) + {Q^6}(174{s^2} - 325su - 209{u^2}) + {Q^4}(126{s^3} - 717{s^2}u\\
\notag
&& + 132s{u^2} - 545{u^3}) + {Q^2}(19{s^4} - 178{s^3}u - 90{s^2}{u^2} - 915s{u^3} - 102{u^4}) - {(s - u)^2}(9{s^3}\\
\notag
&& - 87{s^2}u - 253s{u^2} - 140{u^3})] - 3{Q^{12}} - {Q^{10}}(51s + 32u) + {Q^8}( - 115{s^2} + 189su - 7{u^2})\\
\notag
&& + {Q^6}( - 75{s^3} + 115{s^2}u - 290s{u^2} + 242{u^3}) + 2{Q^4}u( - 192{s^3} + 18{s^2}u + 6s{u^2} + 179{u^3})\\
\notag
&& + 2{Q^2}({s^5} - 126{s^4}u - 27{s^3}{u^2} - 225{s^2}{u^3} + 325s{u^4} + 52{u^5})\\
&& - 2{(s - u)^3}(3{s^3} - 4{s^2}u - 42s{u^2} - 17{u^3})\}
\end{eqnarray}

\begin{large}
\textbf{Appendix B: Full expressions about the condensate terms $\langle\overline{q}q\rangle,\langle\overline{q}g_{s}\sigma Gq\rangle$ and $\langle\overline{q}q\rangle\langle g^{2}_{s}G^{2}\rangle$ for $J/\psi$ off-shell case.}\label{C}
\end{large}

\begin{eqnarray}
{\Pi ^{\left\langle {\bar qq} \right\rangle (J/\psi )}}&& = \left\langle {\bar qq} \right\rangle \frac{{{m_c}}}{{({p^2} - m_c^2)(p{'^2} - m_c^2)}}\\
\notag
{\Pi ^{\left\langle {\bar q{g_s}\sigma Gq} \right\rangle (J/\psi )}}&& =  - \left\langle {\bar q{g_s}\sigma Gq} \right\rangle [\frac{{{m_c}}}{{4{{({p^2} - m_c^2)}^2}(p{'^2} - m_c^2)}}\\
&& + \frac{{2m_c^4}}{{({p^2} - m_c^2){{(p{'^2} - m_c^2)}^3}}} + \frac{{{m_c}}}{{2({p^2} - m_c^2){{(p{'^2} - m_c^2)}^2}}}]\\
\notag
{\Pi ^{\left\langle {\bar qq} \right\rangle \left\langle {g_s^2{G^2}} \right\rangle (J/\psi )}}&& = \left\langle {\bar qq} \right\rangle \left\langle {g_s^2{G^2}} \right\rangle [\frac{{m_c^3}}{{6({p^2} - m_c^2){{(p{'^2} - m_c^2)}^4}}}\\
\notag
&& + \frac{{m_c^5}}{{6({p^2} - m_c^2){{(p{'^2} - m_c^2)}^5}}} + \frac{{m_c^3}}{{12{{({p^2} - m_c^2)}^4}(p{'^2} - m_c^2)}}\\
&& + \frac{{{m_c}}}{{12({p^2} - m_c^2){{(p{'^2} - m_c^2)}^3}}} + \frac{{{m_c}}}{{24{{({p^2} - m_c^2)}^2}{{(p{'^2} - m_c^2)}^2}}}]
\end{eqnarray}

\begin{large}
\textbf{Appendix C: The pole and continuum contributions, and the contributions of different condensate terms for all vertices.}\label{A}
\end{large}

\begin{figure*}[htbp]
\centering
\includegraphics[width=17.5cm]{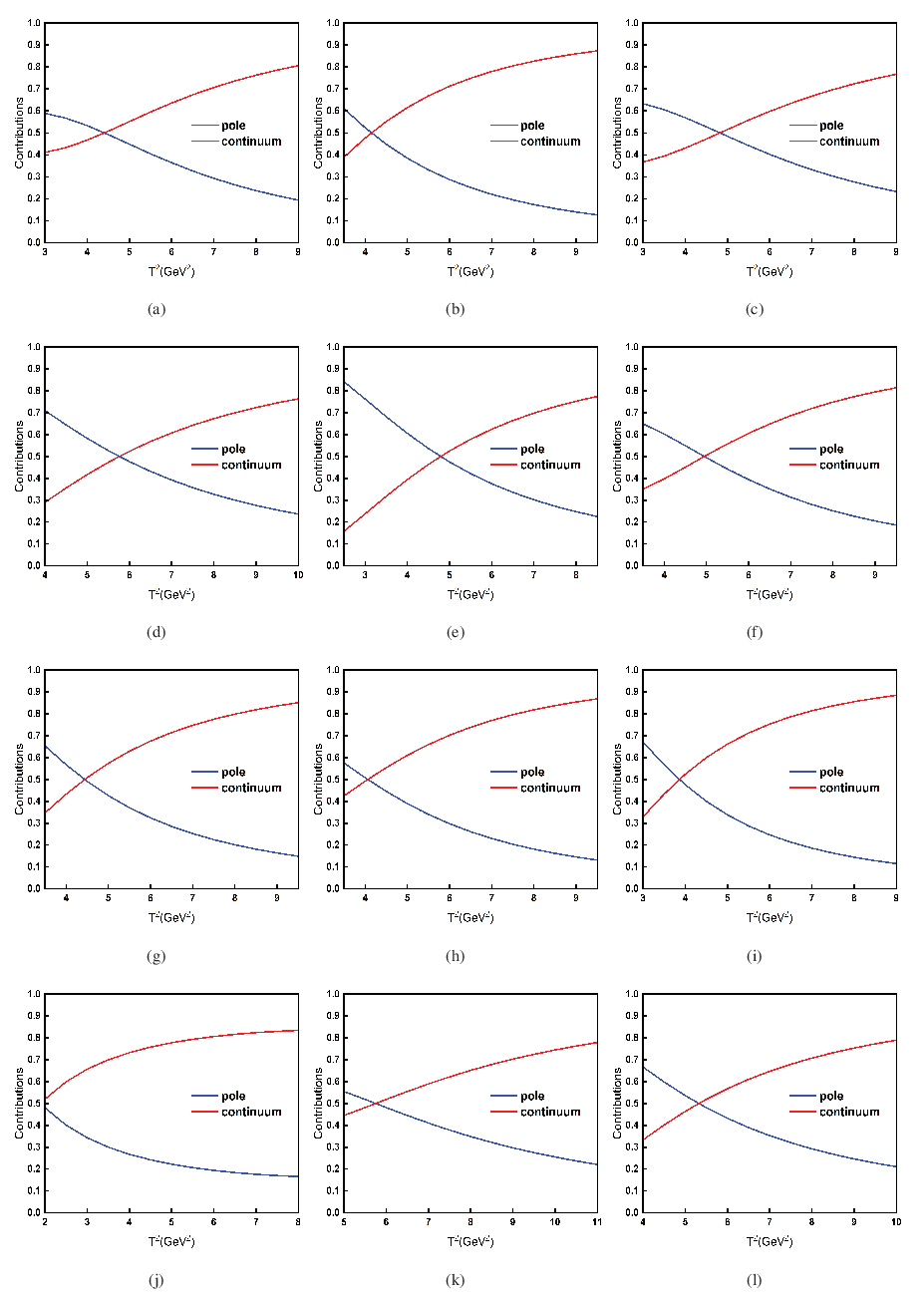}
\caption{The pole and continuum contributions with variation of the Borel parameter $T^{2}$ for $DDJ/\psi$ (a-b), $DD^{*}J/\psi$ (c-e), $D^{*}D^{*}J/\psi$ (f-g), $DD^{*}\eta_{c}$ (h-j), $D^{*}D^{*}\eta_{c}$ (k-l).}
\label{PC}
\end{figure*}

\begin{figure*}[htbp]
\centering
\includegraphics[width=17.5cm]{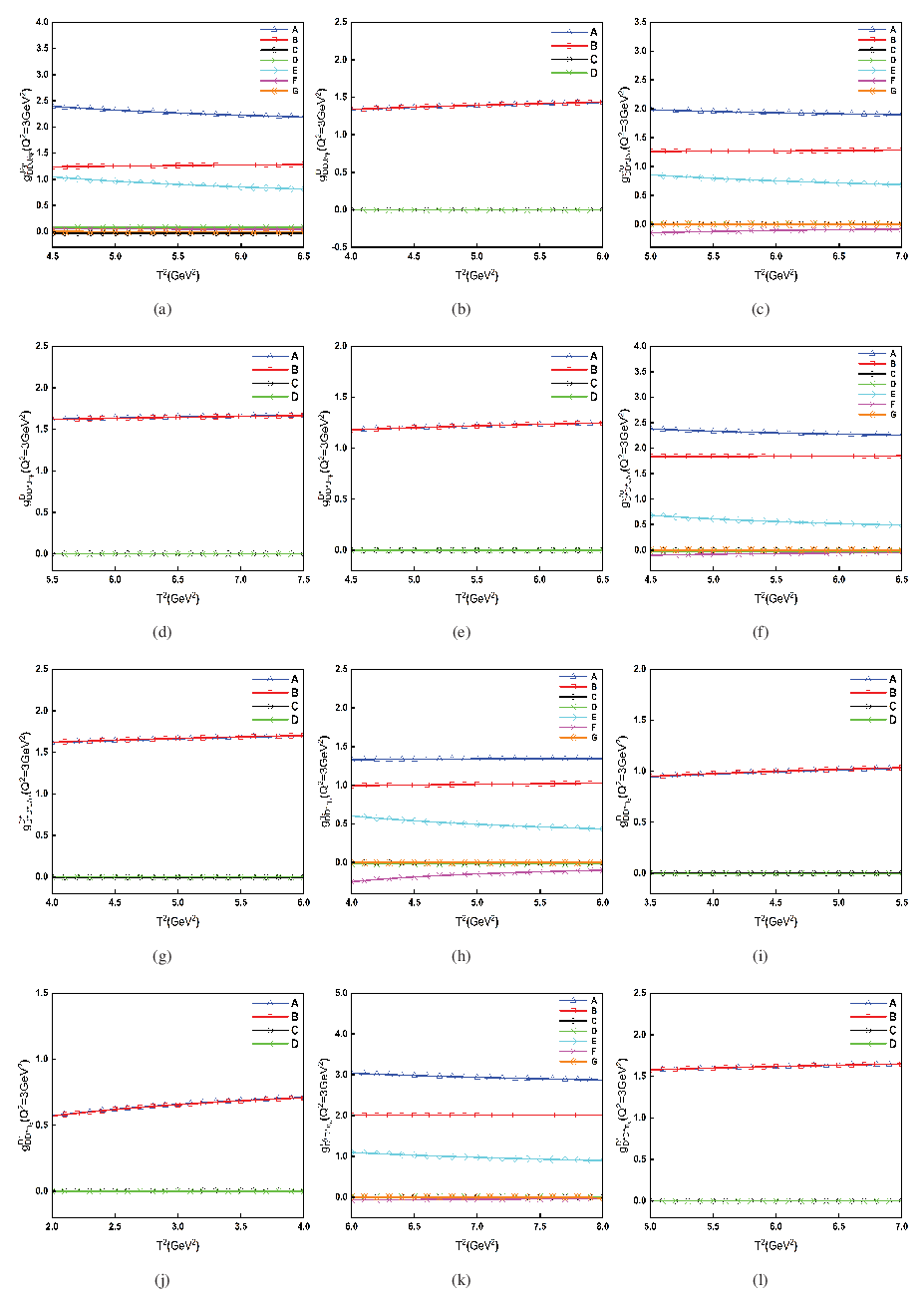}
\caption{The contributions of different vacuum condensate terms with variation of the Borel parameter $T^{2}$ for $DDJ/\psi$ (a-b), $DD^{*}J/\psi$ (c-e), $D^{*}D^{*}J/\psi$ (f-g), $DD^{*}\eta_{c}$ (h-j), $D^{*}D^{*}\eta_{c}$ (k-l), where A-G denote the total, perturbative term, $\langle g_{s}^{2}G^{2}\rangle,\langle f^{3}G^{3}\rangle,\langle\overline{q}q\rangle,\langle\overline{q}g_{s}\sigma G q\rangle$, and $\langle\overline{q}q\rangle\langle g_{s}^{2}G^{2}\rangle$ contributions.}
\label{BW}
\end{figure*}
\end{widetext}

\end{document}